\documentstyle[12pt]{article}
\begin{document}
\def\thefootnote{\fnsymbol{footnote}}
\begin{flushright}
KANAZAWA-06-08  \\
June, 2006
\end{flushright}
\vspace*{2cm}
\begin{center}
{\LARGE\bf Supersymmetric extra U(1) models with a singlino dominated LSP}\\
\vspace{1 cm}
{\Large Satoshi Nakamura}
\footnote[1]{e-mail:~nakamura@hep.s.kanazawa-u.ac.jp}
{\Large and Daijiro Suematsu}
\footnote[2]{e-mail:~suematsu@hep.s.kanazawa-u.ac.jp}
\vspace {1cm}\\
{\it Institute for Theoretical Physics, Kanazawa University,\\
        Kanazawa 920-1192, Japan}\\
\end{center}
\vspace{1cm}
{\Large\bf Abstract}\\
We investigate phenomenology related to the neutral fields 
in supersymmetric models with an extra U(1) derived from $E_6$. 
Our study is concentrated into the models which have a singlino 
dominated neutralino as the lightest superparticle (LSP).
If such models satisfy a constraint for dark matter derived from the WMAP
data, the lightest neutral Higgs scalar, a new neutral gauge field 
$Z^\prime$ and the LSP may be interesting targets for the study 
at the LHC.   
We also discuss features of the $Z^\prime$ in the models and its
detectability at the LHC.

\newpage
\setcounter{footnote}{0}
\def\thefootnote{\arabic{footnote}}
\section{Introduction}
Recent various astrophysical observations quantitatively 
show the existence of a 
substantial amount of non-relativistic and non-baryonic 
dark matter \cite{cmb,sdss}. 
Although supersymmetric models have been considered to
be the best candidate beyond the standard model (SM) 
from a viewpoint of both the gauge hierarchy problem and 
the gauge coupling unification, 
this fact seems to make them much more promising on the basis of an
experimental signature \cite{susy}. 
If $R$-parity is conserved in supersymmetric models, 
the lightest superparticle (LSP) is stable.
Thus the LSP can be a good candidate for cold dark matter (CDM) as long
as it is electrically neutral. 
Since the strength of interactions of the LSP with the SM fields is
 $O(G_F)$ and the mass can be of the order of the weak scale, 
its relic energy density is expected to be eventually of 
the order of critical energy density of the universe. 
The most promising one among such LSP candidates is considered 
to be the lightest neutralino.
Relic abundance of the lightest neutralino has been extensively studied in the 
minimal supersymmetric SM (MSSM) \cite{mssmcdm1,mssmcdm2,mssmcdm3}.
After publication of the analysis of the WMAP data, however, 
the allowed parameter space in the minimal SUGRA (or CMSSM) is found 
to be restricted into some narrow regions \cite{mssmcdm4}.
If we take account of this situation, it seems worth studying
the relic abundance of CDM candidates quantitatively also 
in various extensions of the MSSM.
It may also be useful to discuss indications of such models which are
expected to be found at the forthcoming Large Hadron Collider (LHC) by
applying the CDM condition.

The MSSM has been considered as the most promising supersymmetric 
model and has extensively studied from various points of view.
Although the MSSM can explain experimental results obtained by now
as long as parameters are suitably chosen, 
it suffers from the well known $\mu$ problem \cite{mu}. 
If we try to solve it near the
weak scale, we need to extend the MSSM in the way to give some influence
to physics at TeV regions \cite{nmssm,extra}.
If we extend it by introducing an extra U(1) gauge symmetry, for
example, the problem can be solved in a very elegant way \cite{extra,extra1}.
The existence of additional U(1) symmetries is also predicted in some 
effective theories of superstring \cite{efstring,lhcevid}.
If there is an extra U(1) symmetry at TeV regions, the model is expected 
to reveal distinguishable features from those of the MSSM.
Some signals of the models may be detected at the 
LHC \cite{lhcevid,lhcevid1,gkk} and
a CDM candidate may be different from that in other models.
In this paper we focus our attention into such aspects 
in the models with an extra U(1).

In the models with an extra U(1), an operator $\lambda \hat S\hat H_1\hat H_2$ 
is introduced into superpotential as a gauge invariant operator 
instead of the so-called $\mu$ term 
$\mu \hat H_1\hat H_2$ \cite{extra}.\footnote{In this paper we put
a hat on the character for a superfield. For its component fields, we
put a tilde on the same character to represent the superpartners 
of the SM fields 
and use just the same character without the hat for the SM fields.
Otherwise, the field with no tilde should be understood as a scalar
component.}
In the simplest models to accommodate such a feature, the 
extra U(1) symmetry is supposed to be broken by a vacuum 
expectation value (VEV) of the 
scalar component $S$ of an SM singlet chiral superfield $\hat S$.
The $\mu$ term is generated as $\mu=\lambda\langle S\rangle$ 
by the same singlet scalar field through the introduced operator.
These models show a difference from the MSSM in the neural Higgs
sector in addition to the existence of a new neutral gauge field
$Z^\prime$ \cite{extra,higgsb}.
Since the neutralino sector is extended from the MSSM 
by a fermionic component $\tilde S$ and an extra U(1) 
gaugino $\tilde\lambda_x$, the feature of neutralinos can also
be different from that in the MSSM.
Various phenomena are influenced by this change \cite{sneut}. 

In particular, if the singlino $\tilde S$ can dominate the 
lightest neutralino, distinguishable features from the MSSM 
are expected to appear in the phenomena relevant to the neutralinos.
When $\langle S\rangle$ takes a value of the weak scale, the singlino
domination of the lightest neutralino is naturally expected to occur. 
This is the case in the well known next MSSM (NMSSM) \cite{nmssm0} and
its modified model (nMSSM) \cite{nmssm1}.
In the models with an extra U(1), however, there are severe constraints 
on $\langle S\rangle$ from both mass bounds of $Z^\prime$ which 
result from the direct search of $Z^\prime$ \cite{cdf}
and bounds on the mixing between $Z^\prime$ and the ordinary $Z$ 
which result from the electroweak precision measurements 
\cite{lhcevid,pdg,kmix1}.
These constraints tend to require that $\langle S\rangle$ should be
more than $O(1)$~TeV as long as we do not consider a special
situation.\footnote{Even in the models with an extra U(1), if one considers 
a model with a secluded singlet sector which is called the $S$-model 
in \cite{secl}, $\langle S\rangle$ can take a value of the weak scale. 
In this case phenomenological feature at the weak scale is very 
similar to the nMSSM.} 
Thus, in the simple models with an extra U(1) which is called the UMSSM
in \cite{secl}, it seems unable to expect 
substantial differences in the lightest neutralino from 
the MSSM since both $\tilde S$ and $\tilde\lambda_x$ tend to decouple
from the lightest neutralino. 

This situation changes if the extra U(1) gaugino $\tilde\lambda_x$ is
sufficiently heavier than $Z^\prime$ as suggested in 
\cite{extra1,ce,secl,singlino}. 
In these papers the nature of the lightest neutralino has been studied 
to show a lot of interesting aspects:  
(a)~it can be dominated by the singlino $\tilde S$ and can be very light;
(b)~it can be a good candidate of the CDM whose nature is very different 
from that in the MSSM. It can have a small mass which has already 
been forbidden in the MSSM since it is dominated by the singlino;
(c)~it also has a different interaction from that in the MSSM. 
However, the studied parameter regions are different in each study.
In \cite{ce} the CDM abundance is mainly studied under the assumption
that the lightest neutral Higgs mass is $m_h=170$~GeV. 
In the S-model case the neutralino mass 
matrix is assumed to be reduced to the one of the nMSSM \cite{secl} or 
a certain gaugino mass relation such as $M_{\tilde x}=M_{\tilde Y}$ 
is assumed \cite{bll}.
In this paper we do not use these assumptions. We are interested in
other region of the parameter space, where
$\tilde\lambda_x$ is much heavier than other gauginos
but it has non-negligible mixing with $\tilde S$ \cite{singlino,mixing}. 
Therefore $\tilde\lambda_x$ can give a crucial contribution to the mass
of $\tilde S$.
We focus our study on such cases and investigate the neutral field 
sector in the models with an extra U(1) derived from $E_6$. 
We discuss phenomenological features of the
models and study indications of the models expected to be found at
the LHC.  

The paper is organized as follows. In section 2 we define the models and
discuss their features different from other models.  
In particular, we focus our attention on the neutral field sectors 
of the models, that is, 
the lightest neutralino, the new neutral gauge field $Z^\prime$ 
and the neutral Higgs scalars.
In section 3 we study the parameter space of the models which is allowed
by the current results of various experiments. We predict signals of 
the models expected to be seen at the LHC. 
Section 4 is devoted to the summary. 
 
\section{$\mu$ problem solvable extra U(1) models} 
\subsection{Features of the neutral fields}
In the simple models with an extra U(1) which can solve the $\mu$
 problem, the mass of $Z^\prime$ is directly related to the $\mu$ 
term \cite{extra,extra1}. 
This feature induces various interesting phenomena which make the models
distinguishable from the MSSM. 
The $\mu$ term is considered to be generated 
by an operator in the last term of the superpotential 
\begin{equation}
W_{\rm ob}=h_U\hat{\bar U}\hat Q\hat H_2+h_D\hat{\bar D}\hat Q\hat H_1
+h_E\hat{\bar E}\hat L\hat H_1+\lambda \hat S\hat H_1\hat H_2,
\label{mu}
\end{equation}
where $\hat H_{1,2}$ are the ordinary doublet Higgs chiral superfields 
and $\hat S$ is an additional singlet chiral superfield.
If a VEV of the scalar component of $\hat S$
is assumed to generate both the $\mu$ term 
and the $Z^\prime$ mass, the superpotential (\ref{mu}) requires 
that $\hat H_{1,2}$ also have the extra U(1) charges $Q_{1,2}$.
The charge conservation imposes a condition $Q_1+Q_2+Q_S=0$ on them.
A bare $\mu$ term $\mu \hat H_1\hat H_2$ is 
automatically forbidden by this symmetry.
Stability of the scalar potential for $S$ is automatically 
guaranteed by a quartic term induced as the extra U(1) $D$-term without
introducing a new term in the superpotential. 
The $Z^\prime$ mass is difficult to be much larger 
than $O(1)$~TeV as long as there is no other contribution to it.
These aspects may make the models not only theoretically interesting 
but also a good target for the studies at the LHC \cite{lhcevid,lhcevid1,gkk}. 
In particular, we can find interesting features in the neutral 
fields, which make the models distinguishable from the MSSM and also the 
singlet extensions of the MSSM such as the NMSSM, the nMSSM and the S-model.
In this section we review these features to clarify the differences of
our models from previously studied ones.

Before proceeding to the discussion on this issue, we fix assumptions for the
supersymmetry breaking.
We assume soft supersymmetry breaking terms
\begin{eqnarray}
-{\cal L}_{\rm SUSY}&=&\sum_\varphi m_\varphi^2|\varphi|^2 + 
\left({1\over 2}M_{\tilde g}\tilde\lambda_g\tilde\lambda_g
+{1\over 2}M_{\tilde W}\tilde\lambda_W\tilde\lambda_W
+{1\over 2}M_{\tilde Y}\tilde\lambda_Y\tilde\lambda_Y
+{1\over 2}M_{\tilde x}\tilde\lambda_x\tilde\lambda_x 
+{\rm h.c.}\right) \nonumber\\
&-&\left(A_Uh_U\hat{\bar U}\hat Q\hat H_2
+A_Dh_D\hat{\bar D}\hat Q\hat H_1
+A_Eh_E\hat{\bar E}\hat L\hat H_1+A_\lambda\lambda \hat S\hat H_1\hat H_2
+{\rm h.c.}\right),
\label{soft}
\end{eqnarray}
where $\varphi$ in the first term runs all scalar fields 
contained in the models.
Scalar mass $m_\varphi$ and $A$ parameters for scalar 
trilinear terms may be assumed to have a universal value $m_{3/2}$. 
A grand unification relation for the
masses of gauginos may also be assumed.   

Now we assume that the scalar components of $\hat H_{1,2}$ and $\hat S$ obtain 
the VEVs $v_1$, $v_2$ and $u$ due to radiative effects on the supersymmetry
breaking parameters and the electroweak symmetry 
SU(2)$_L\times$U(1)$_Y\times$U(1)$_x$ breaks down 
into U(1)$_{\rm em}$ \cite{extra,extra1}.
We define the field fluctuations around this vacuum as
\begin{equation}
H_1=\left(\begin{array}{c}v_1+h_1^0+iP_1 
\\ h_1^-\\ \end{array}\right),
\quad
H_2=\left(\begin{array}{c}h_2^+ 
\\ v_2+h_2^0+iP_2\\\end{array}\right),
\quad
S=u+h_S^0+iP_S.
\label{vev}
\end{equation}
The last term in eq.~(\ref{mu}) generates a $\mu$ parameter as 
$\mu=\lambda u$ and plays a required role for the $\mu$ term in the MSSM. 
Although the models have similar structures to the MSSM after this symmetry
breaking, there appear various differences induced by this term, in
particular, in the sector of the neutral fields.
In a charged field sector, we can find a difference from the MSSM
only in the mass of the charged Higgs scalar. It will be discussed 
in Appendix.

\subsubsection{Neutral gauge field sector}
The clearest difference from the MSSM is the existence of the $Z^\prime$
at TeV regions.
Through the symmetry breaking denoted by eq.~(\ref{vev}), 
a mass matrix $M_{ZZ^\prime}^2$ is generated for the neutral 
gauge bosons $Z_\mu$ and $Z^\prime_\mu$. It can be written in the basis 
$(Z_\mu, Z_\mu^\prime)$ as
\begin{equation}
\left(\begin{array}{cc}
{g_2^2+g_1^2\over 2}v^2 & 
{g_x\sqrt{g_2^2+g_1^2}\over 2}v^2(Q_1\cos^2\beta-Q_2\sin^2\beta) \\
{g_x\sqrt{g_2^2+g_1^2}\over 2}v^2(Q_1\cos^2\beta-Q_2\sin^2\beta) &
{g_x^2\over 2}v^2(Q_1^2\cos^2\beta+Q_2^2\sin^2\beta+Q_S^2{u^2\over v^2}) \\
\end{array}\right)
\label{zmass}
\end{equation}
where $v^2=v_1^2+v_2^2$ and $\tan\beta=v_2/v_1$.
The extra U(1) charge $Q_x(f)$ of a field $f$ and its coupling $g_x$ are 
defined in such a way that the covariant derivative for the relevant
gauge group
SU(2)$_L\times$U(1)$_Y\times$U(1)$_x$ takes the form
\begin{equation}
D_\mu=\partial_\mu-ig_2{\tau^a\over 2}W_\mu^a-ig_1{Y(f)\over 2}B_\mu
-ig_x{Q_x(f)\over 2}Z_\mu^\prime.
\label{cov}
\end{equation}
Mass eigenvalues of the neutral gauge fields $Z_{1\mu}$ and $Z_{2\mu}$
can be approximately expressed as
\begin{eqnarray}
&&m_{Z_1}^2\simeq m_Z^2-m_Z^2{g_x\tan 2\xi\over\sqrt{g_2^2+g_1^2}}
(Q_1\cos^2\beta-Q_2\sin^2\beta), \nonumber \\ 
&&m_{Z_2}^2\simeq
{g_x^2\over 2}(Q_1^2v_1^2+Q_2^2v_2^2+Q_S^2u^2)
+m_Z^2{g_x\tan 2\xi\over\sqrt{g_2^2+g_1^2}}
(Q_1\cos^2\beta-Q_2\sin^2\beta), 
\label{z2}
\end{eqnarray}
where $m_Z$ is the mass of the $Z$ boson in the SM and $\xi$ is a $ZZ^\prime$ 
mixing angle defined by
\begin{equation}
\tan 2\xi={2g_x\sqrt{g_2^2+g_1^2}(Q_1\cos^2\beta-Q_2\sin^2\beta)
\over g_x^2(Q_1^2\cos^2\beta+Q_2^2\sin^2\beta+Q_S^2u^2/v^2)-(g_2^2+g_1^2)}.
\label{mix}
\end{equation}
If $u\gg v_{1,2}$ is satisfied, $m_{Z_1}$ approaches to $m_Z$ and 
$m_{Z_2}$ is proportional to $u$. 
The relation between $m_{Z_2}$ and $u$ is given by 
$m_{Z_2}/u\simeq g_xQ_S/\sqrt{2}$. 

Both direct search of a new neutral gauge field and precise
measurements of the electroweak interaction severely constrain the
mass eigenvalue $m_{Z_2}$ of the new neutral gauge boson and the $ZZ^\prime$ 
mixing angle $\xi$ \cite{pdg}. 
Lower bounds for $m_{Z_2}$ have been studied by using the
searches of the $Z_2$ decay into dilepton pairs \cite{cdf}.
Although it depends on the models, it may be roughly estimated as
$m_{Z_2}~{^>_\sim}~600$~GeV.
If $Z_2$ has a substantial decay width into non-SM fermion pairs 
such as neutralino pairs, this bound may be largely relaxed \cite{gkk}.
On the other hand, the precise measurements of the electroweak
interaction give a constraint $\xi~{^<_\sim}~10^{-3}$ \cite{pdg}.
As found from eq.~(\ref{mix}), this bound can be fulfilled if
either of two conditions is satisfied, that is, a sufficiently large $u$
or $\tan\beta\simeq\sqrt{Q_1/Q_2}$ \cite{extra1}. 
For the latter case, since the constraint from the $ZZ^\prime$ mixing 
can automatically be guaranteed, $u$ needs not so large 
as long as the direct search constraint on $m_{Z_2}$ is satisfied.
This seems to be an important point to be noted when we consider 
the existence of an extra U(1) at TeV regions. We will focus our study on
this case, which may make it possible to find the solutions in the
parameter regions excluded in the study of the UMSSM \cite{ce,secl}.

\subsubsection{Neutralino sector}
The neutralino sector is extended into six components, 
since there are two additional neutral fermions 
$\tilde\lambda_x$ and $\tilde S$. $\tilde\lambda_x$ is the extra
U(1) gaugino and $\tilde S$ is the fermionic component of $\hat S$.  
If we take a basis 
${\cal N}^T=(-i\tilde\lambda_x, -i\tilde\lambda_W^3, -i\tilde\lambda_Y, 
\tilde H_1, \tilde H_2, \tilde S)$ and define
neutralino mass terms such as
${\cal L}_{\rm neutralino}^m=-{1\over 2}{\cal N}^T{\cal MN}+{\rm h.c.}$,
a 6 $\times$ 6 neutralino mass matrix ${\cal M}$ can be represented
as\footnote{We do not consider gauge kinetic term mixing 
between U(1)$_Y$ and the extra U(1), for simplicity. 
The study of their phenomenological effects can be found in \cite{sneut}.} 
\begin{equation}
\left( \begin{array}{cccccc}
M_{\tilde x} & 0 & 0 & {g_xQ_1\over \sqrt 2}v\cos\beta 
& {g_xQ_2\over \sqrt 2}v\sin\beta & {g_xQ_S\over \sqrt 2}u \\
0 & M_{\tilde W} & 0 &m_Zc_W\cos\beta & -m_Zc_W\sin\beta &0 \\
0 & 0 & M_{\tilde Y} & -m_Zs_W\cos\beta & m_Zs_W\sin\beta &0 \\
{g_xQ_1\over \sqrt 2}v\cos\beta &m_Zc_W\cos\beta &-m_Zs_W\cos\beta & 0 & 
\lambda u & \lambda v\sin\beta \\
{g_xQ_2\over \sqrt 2}v\sin\beta & -m_Zc_W\sin\beta & m_Zs_W\sin\beta 
& \lambda u & 0 & \lambda v\cos\beta \\
{g_xQ_S\over \sqrt 2}u & 0 & 0 & \lambda v\sin\beta & \lambda v\cos\beta & 0\\
\end{array} \right).
\end{equation}
Neutralino mass eigenstates $\tilde\chi_a^0(a=1\sim 6)$ are related 
to ${\cal N}_j$ by using the mixing matrix $U$ as
\begin{equation}
\tilde\chi^0_a=\sum_{j=1}^6U_{aj}{\cal N}_j,
\label{meig}
\end{equation}
where $U$ is defined in such a way that $U{\cal M}U^T$ becomes diagonal.

The composition of the lightest neutralino is important for the study of 
various phenomena, in particular, the relic density of the
lightest neutralino as a CDM candidate. 
If $u$ is a similar order value to $v_{1,2}$ 
or less than these, the lightest neutralino is expected to be 
dominated by the singlino $\tilde S$ just like in the case of the
nMSSM and the S-model. In this case, if it can annihilate sufficiently, 
the lightest neutralino with a sizable singlino component may be 
a good CDM candidate in the parameter regions different
from the ones in the MSSM \cite{nmssm0,nmssm1}.
In the present models, however, the $Z^\prime$ constraints seem to 
require that $u$ should be much larger than $v_{1,2}$ as mentioned before. 
As the result, $\tilde\lambda_x$ and $\tilde S$ tend to decouple from 
the lightest neutralino as long as the mass of $\tilde\lambda_x$ 
is assumed to be similar to the masses of other gauginos.
In such a situation, the composition of the lightest neutralino 
is expected to be similar to that of 
the MSSM. Then we cannot find distinctive features in the lightest 
neutralino.
However, as suggested in \cite{extra1,ce,secl,singlino}, 
this situation can be drastically changed if the mass of the extra 
U(1) gaugino $M_{\tilde x}$ becomes 
much larger than the masses of other gauginos due to some reasons. 
In this case the lightest neutralino can be dominated by 
the singlino $\tilde S$.

If the gaugino $\tilde\lambda_x$ is heavy enough to satisfy 
$M_{\tilde x} \gg {g_xQ_S\over \sqrt 2}u$, we can integrate out 
$\tilde\lambda_x$ just as the seesaw mechanism for the neutrinos. 
A resulting $5\times 5$ mass matrix can be expressed as
\begin{equation}
\left( \begin{array}{ccccc}
M_{\tilde W} & 0 &m_Zc_W\cos\beta & -m_Zc_W\sin\beta &0 \\
0 & M_{\tilde Y} & -m_Zs_W\cos\beta & m_Zs_W\sin\beta &0 \\
m_Zc_W\cos\beta &-m_Zs_W\cos\beta &
-{g_x^2Q_1^2\over 2M_{\tilde x}}v^2\cos^2\beta   & 
\lambda u & \lambda v\sin\beta \\
-m_Zc_W\sin\beta & m_Zs_W\sin\beta & \lambda u & 
-{g_x^2Q_2^2\over 2M_{\tilde x}}v^2\sin^2\beta & \lambda v\cos\beta \\
 0 & 0 & \lambda v\sin\beta & \lambda v\cos\beta & 
-{g_x^2Q_S^2\over 2M_{\tilde x}}u^2\\
\end{array} \right).
\label{mchi}
\end{equation}
This effective mass matrix suggests that the lightest neutralino 
tends to be dominated by the singlino $\tilde S$ 
as long as $M_{\tilde W,\tilde Y}$ and $\mu(\equiv\lambda u)$ 
is not smaller than
${g_x^2Q_S^2u^2\over 2M_{\tilde x}}$. Since $M_{\tilde W}$ 
and $\mu$ cannot to be
less than $100$~GeV because of mass bounds of the lightest chargino and
the gluino \cite{chargino}, 
the singlino domination of the lightest neutralino is expected 
to be naturally realized in the case that $M_{\tilde x}\gg u$
is satisfied. 
In such a case, phenomenology of the lightest neutralino can be
largely changed from that in the MSSM and also its singlet extensions.
This effective mass matrix reduces to the one of the nMSSM
in the large $M_{\tilde x}$ limit \cite{secl}.
However, we are interested in the intermediate situation where the effectively
generated diagonal elements in eq.~(\ref{mchi}) can not be neglected.
Since this possibility has not been studied in detail in realistic models
yet despite it is potentially interesting,\footnote{
In particular, the detailed study seems to have not been done 
under the assumption $\tan\beta=\sqrt{Q_1/Q_2}$. 
In \cite{singlino} we do not consider the anomaly problem of U(1)$_x$ 
seriously and only study a toy model.} 
we will concentrate our study into such a situation in this paper.
The lightest neutralino is assumed to be dominated by the singlino 
because of a large $M_{\tilde x}$ compared with the mass of 
other gauginos. 

\subsubsection{Neutral Higgs scalar sector}
The neutral Higgs mass is also modified from that 
in the MSSM as in the case of the NMSSM and the nMSSM.
A difference from the latter ones is the existence of an additional 
$D$-term contribution of the extra U(1). 
The CP even neutral Higgs sector is composed of the three scalars
$(h_1^0, h_2^0, h_S^0)$ which are introduced in eq.~(\ref{vev}). 
Their mass matrix ${\cal M}_h^2$ at tree level is written as
\begin{equation}
\footnotesize
\left(\begin{array}{ccc}
{1\over 2}(g_2^2+g_1^2+g_x^2Q_1^2)v_1^2+A_\lambda\lambda u\tan\beta &
-{1\over 2}(g_2^2+g_1^2-g^2(1,2))v_1v_2-A_\lambda\lambda u & 
{1\over 2}g^2(1,S)v_1u-A_\lambda\lambda v_2 \\
-{1\over 2}(g_2^2+g_1^2-g^2(1,2))v_1v_2-A_\lambda\lambda u & 
{1\over 2}(g_2^2+g_1^2+g_x^2Q_2^2)v_2^2+A_\lambda\lambda u\cot\beta &
{1\over 2}g^2(2,S)v_2u-A_\lambda\lambda v_1 \\
{1\over 2}g^2(1,S)v_1u-A_\lambda\lambda v_2 &
{1\over 2}g^2(2,S)v_2u-A_\lambda\lambda v_1 &
{1\over 2}g_x^2Q_S^2u^2+A_\lambda\lambda {v_1v_2\over u} \\
\end{array}\right),
\normalsize
\label{hmass}
\end{equation}
where $A_\lambda$ is the soft supersymmetry breaking parameter 
defined in eq.~(\ref{soft}).
We use a definition $g^2(i,j)=g_x^2Q_iQ_j+4\lambda^2$ in this formula.
Mass eigenstates $\phi_\alpha$ are related to the original neutral 
CP even Higgs scalars $h_a^0$ by
\begin{equation}
\phi_\alpha=\sum_{a=1,2,S}{\cal O}_{\alpha a}h_a^0,
\label{hmix}
\end{equation}
where the orthogonal matrix ${\cal O}$ is defined so as to diagonalize
the neutral Higgs mass matrix ${\cal M}_h^2$ in such a way as
${\cal O}{\cal M}_h^2{\cal O}^T
={\rm diag}(m_{\phi_1}^2,m_{\phi_2}^2,m_{\phi_3}^2)$.

Since upper bounds of the mass eigenvalue for the lightest 
neutral Higgs scalar $h^0$ can be 
estimated by using eq.(\ref{hmass}) as  
\begin{equation}
m_{h^0}^2\le m_Z^2\left[\cos^22\beta
+{2\lambda^2\over g_2^2+g_1^2}\sin^22\beta+
{g_x^2\over g_2^2+g_1^2}(Q_1\cos^2\beta+Q_2\sin^2\beta)^2\right] 
+\Delta m_1^2,
\label{higgs}
\end{equation}
it can be larger than that in the MSSM. 
The second term in the brackets of eq.~(\ref{higgs}) comes 
from the interaction given by the last term in eq.~(\ref{mu}). 
It can give a large contribution for smaller values of $\tan\beta$ 
and $u$ for a fixed $\mu$.
The third term is the $D$-term contribution of the extra U(1).
Due to these effects, even in the regions of the small $\tan\beta$ such
as $\tan\beta= 1-2$, the mass of
the lightest neutral Higgs $m_{\phi_\alpha}$ can take larger values 
than that in the
MSSM, such as 120 GeV or more, if one-loop corrections 
$\Delta m_1^2$ are taken into account \cite{higgsb}. 
Since dominant components of this lightest Higgs scalar are 
expected to be $h_{1,2}^0$ as long as $u$ is not smaller than
$v_{1,2}$, its nature is similar to that in the MSSM, except that
it is heavier than that in the MSSM. This situation 
seems very different from the nMSSM and the S-model where the 
lightest neutral Higgs is a mixture state of $h_{1,2}^0$ and $h_S^0$.
In numerical studies given in section 3, we will estimate 
both the Higgs mass eigenvalues and their eigenstates by
diagonalizing the mass matrix (\ref{hmass}) including the one-loop 
corrections due to the stops.  

CP odd Higgs scalars are also somewhat changed from the ones in the MSSM.
A CP odd Higgs mass matrix ${\cal M}_P^2$ can be written as  
\normalsize
\begin{equation}
{\cal M}_P^2=\left(\begin{array}{ccc}
A_\lambda\lambda u\tan\beta & A_\lambda\lambda u & A_\lambda\lambda v_2 \\
A_\lambda\lambda u & A_\lambda\lambda u\cot\beta & A_\lambda\lambda v_1 \\
A_\lambda\lambda v_2 & A_\lambda\lambda v_1 & 
A_\lambda\lambda {v_1v_2\over u}\\
\end{array}\right).
\end{equation}
Only one component $P_A$ has a non-zero mass eigenvalue
\begin{equation}
m^2_{P_A}={2A_\lambda\lambda u\over\sin 2\beta}
\left(1+{v^2\over 4u^2}\sin^22\beta\right),
\label{mpseudo}
\end{equation}
and others are would-be Goldstone bosons $G_{1,2}^0$ as in the MSSM. 
This requires $\lambda uA_\lambda>0$ for the stability of the vacuum.
Imaginary parts of the original Higgs fields in eq.~(\ref{vev}) 
have $P_A$ as a component. They can be written as
\begin{equation}
P_1={u\sin\beta\over N}P_A+\dots, \quad
P_2={u\cos\beta\over N}P_A+\dots, \quad
P_S={v\sin\beta\cos\beta\over N}P_A+\dots,
\label{cpodd}
\end{equation}
where a normalization factor $N$ is defined as 
$N=\sqrt{u^2+v^2\sin^2\beta\cos^2\beta}$.
Although $m^2_{P_A}$ takes larger values than those in the MSSM, 
$P_A$ is found to be similar to that in the MSSM if $u$ becomes larger
than $v$.

\subsection{A CDM constraint}
In the models with a large $M_{\tilde x}$ the composition and the
interaction of the lightest neutralino can 
be very different from that in the MSSM and its singlet extensions. 
Since the lightest neutralino can also be a CDM candidate in such models, 
the relic abundance is expected to give a different constraint 
on the parameter space from that in the MSSM and its singlet extensions. 
It is useful to discuss this point briefly here.

The relic abundance of the stable lightest neutralino
$\tilde\chi^0_\ell$ which is thermally produced can be evaluated as 
thermal abundance at its freeze-out temperature $T_F$.
This temperature can be determined by 
$H(T_F)\sim \Gamma_{\tilde\chi^0_\ell}$.
$H(T_F)$ is the Hubble parameter at $T_F$ \cite{xf}. 
$\Gamma_{\tilde\chi^0_\ell}$ is an annihilation 
rate of $\tilde\chi^0_\ell$ and it can be expressed as
$\Gamma_{\tilde\chi^0_\ell}=\langle\sigma_{\rm ann} v\rangle
n_{\tilde\chi^0_\ell}$, 
where $\langle\sigma_{\rm ann}v\rangle$ is thermal average of 
the product of an annihilation cross section $\sigma_{\rm ann}$
and relative velocity $v$ of annihilating $\tilde\chi^0_\ell$s.
Thermal number density of non-relativistic $\tilde\chi^0_\ell$ 
at this temperature is expressed by $n_{\tilde\chi^0_\ell}$. 
If we use the non-relativistic expansion for the annihilation cross section
such as $\sigma_{\rm ann}v\simeq a+bv^2$ and introduce a dimensionless 
parameter $x_F=m_{\tilde\chi^0_\ell}/T_F$, 
we find that $x_F$ can be represented as
\begin{equation}
x_F=\ln{0.0955m_{\rm pl}m_{\tilde\chi^0_\ell}(a+6b/x_F)
\over (g_\ast x_F)^{1/2}},
\end{equation}  
where $m_{\rm pl}$ is the Planck mass and $g_\ast$ enumerates 
the degrees of freedom of relativistic particles at $T_F$. 
Using this $x_F$, the present abundance of $\tilde\chi^0_\ell$  
can be estimated as
\begin{equation}
\Omega_\chi h^2|_0=
\left.{m_{\tilde\chi^0_\ell} 
n_{\tilde\chi^0_\ell}\over \rho_{\rm cr}/h^2 }\right|_0
\simeq{8.76\times 10^{-11}g_\ast^{-1/2}x_F\over 
(a+3b/x_F)~{\rm GeV}^2 }.
\end{equation} 
We can find formulas of the coefficients $a$ and $b$ for the
processes mediated by the exchange of various
fields contained in the MSSM in the articles \cite{mssmcdm1,mssmcdm2}.

If the lightest neutralino $\tilde\chi^0_\ell$ is dominated by the
singlino and also relatively light, the decay modes into other 
final states than 
the SM fermion-antifermion pairs are expected to be suppressed. 
The singlino dominated neutralino contains the MSSM higgsinos 
(${\cal N}_{4,5}$) and gauginos (${\cal N}_{2,3}$) as its components
with an extremely small ratio.
Thus the annihilation process caused by these components are heavily
suppressed and then cannot be dominant modes unless the enhancement due
to the pole effects of the intermidiate fields. As long as we consider
this lightest neutralino is relatively light, such pole enhancements are
kinematically forbidden in the annihilation modes which have gauge
bosons ($W^+W^-, Z_1Z_1$) and Higgs scalars as the final states.
Then we can expect that the annihilation through the modes 
$\tilde\chi^0_\ell\tilde\chi^0_\ell\rightarrow f\bar f$ is dominant,
since the above mentioned situation is escapable for this process.
Moreover, exotic fields are considered to be heavy enough and then cannot
be the final states kinematically.
Thus $f$ is expected to be restricted to quarks and leptons. 
In the present analyses we will mainly focus our attention to this case.  
These annihilation processes of the lightest neutralino in the models 
with an extra U(1) are expected to be mediated by the exchange 
of $Z_1$, $Z_2$ and the neutral Higgs scalars in the $s$-channel 
and by the sfermion exchange in the $t$-channel. 
New interactions related to these annihilation processes 
of the lightest neutralino $\tilde\chi_\ell^0$ can be
written as
\begin{eqnarray}
{\cal L}&=&\sum_{j=4}^6{g_xQ_j\over 2}
\bar{\cal N}_j\gamma_5\gamma^\mu{\cal N}_j
Z^\prime_\mu +\lambda\left(h^0_1{\cal N}_5{\cal N}_6
+h^0_2{\cal N}_4{\cal N}_6+h^0_3{\cal N}_4{\cal N}_5 \right), \nonumber\\
&+&{g_xQ(f)\over\sqrt 2}\left(\bar{\cal N}_1\bar f\tilde f
-{\cal N}_1f\tilde f^\ast\right)+\dots, \nonumber \\
&\simeq& \sum_{j=4}^6{g_xQ_j\over 2}U_{\ell j}^2
\bar{\tilde\chi}_j^0\gamma_5\gamma_\mu \tilde\chi_j^0Z^\mu_2
+\lambda U_{\ell 6}(U_{\ell 4}{\cal O}_{\alpha 2}
+U_{\ell 5}{\cal O}_{\alpha 1})
\tilde\chi_\ell^0\tilde\chi_\ell^0\phi_\alpha \nonumber \\
&+&{g_xQ(f)\over\sqrt 2}U_{\ell 1}\tilde\chi_\ell^0
\left(\bar f\tilde f -f\tilde f^\ast\right)+ \dots,  
\label{newint}
\end{eqnarray}
where we use eqs.~(\ref{meig}) and (\ref{hmix}).

If we consider the case that the singlino dominates the lightest 
neutralino, the annihilation cross section into the final 
states $f\bar f$ is expected to obtain the dominant contributions from 
the exchange of the new neutral gauge field $Z_2$ and the exchange 
of the lightest neutral Higgs scalar $\phi_\alpha$.
They crucially depend on both the composition and the mass
of $\tilde\chi_\ell^0$ and $\phi_\alpha$. 
These contributions to $a$ and $b$ can be expressed as \cite{ce}
\begin{eqnarray}
a_f&=&{2c_f\over\pi}
\left[{m_fg_x^2\sum_{j=4}^6{Q_j\over 2}U_{\ell j}^2
\over 4m_{\tilde\chi_\ell^0}^2-m_{Z_2}^2}
\left({Q(f_L)\over 2}-{Q(f_R)\over 2}\right)\right]^2
\left({1-{m_f^2\over m_{\tilde\chi^0_\ell}^2}}\right)^{1/2}+\dots, 
\nonumber \\
b_f&=&{1\over 6}\left(-{9\over 2}
+{3 \over 4}{m_f^2\over m_{\tilde\chi^0_\ell}^2-m_f^2}\right)a_f  \nonumber \\ 
&+&{c_f\over 3\pi}  
\left[{m_{\tilde\chi^0_\ell} g_x^2
\sum_{j=4}^6{Q_j\over 2}U_{\ell j}^2
\over 4m_{\tilde\chi^0_\ell}^2 -m_{Z_2}^2}\right]^2 
\left[\left({Q(f_L)\over 2}\right)^2+ \left({Q(f_R)\over 2}\right)^2\right] 
\left(4+{2m_f^2\over m_{\tilde\chi^0_\ell}^2}\right)
\left({1-{m_f^2\over m_{\tilde\chi^0_\ell}^2}}\right)^{1/2} \nonumber \\
&+&{c_f\over 8\pi}
\left({m_{\tilde\chi_\ell^0}\over 4m_{\tilde\chi_\ell^0}^2
-m_{\phi_\alpha}^2 }{\lambda m_f\over v}{\cal P}_f\right)^2
\left({1-{m_f^2\over m_{\tilde\chi^0_\ell}^2}}\right)^{3/2}+\dots, 
\label{zprime}
\end{eqnarray} 
where $c_f=1$ for leptons and 3 for quarks.
The extra U(1) charges of the fermions $f_{L,R}$ are denoted by $Q(f_L)$ 
and $Q(f_R)$. ${\cal P}_f$ is defined by using
eq.~(\ref{cpodd}) as
\begin{equation}
{\cal P}_f=\left\{\begin{array}{ll}
\displaystyle {1\over \sin\beta}
U_{\ell 6}{\cal O}_{\alpha 2}
(U_{\ell 4}{\cal O}_{\alpha 2}+U_{\ell 5}{\cal O}_{\alpha 1}) & 
(f~{\rm with}~T_3={1\over 2})  \\
\displaystyle {1\over \cos\beta}
U_{\ell 6}{\cal O}_{\alpha 1}
(U_{\ell 4}{\cal O}_{\alpha 1}+U_{\ell 5}{\cal O}_{\alpha 2}) & 
(f~{\rm with}~T_3=-{1\over 2})  \\
\end{array}\right.,
\end{equation} 
where $T_3$ is the weak isospin. Contributions due to the CP odd and
heavier CP even Higgs scalars are represented by the ellipses in
eq.~(\ref{zprime}). They are expected to be suppressed because of their
large masses.
  
Since the second term of $b_f$ has no suppression from the masses of the
final state fermions, all quarks and leptons can contribute to this term
as long as the threshold is opened. 
This contribution can be effective for a larger 
$m_{\tilde\chi^0_\ell}(<m_{Z_2})$
even in the case that $a_f$ is suppressed by a large value of $m_{Z_2}$.
The third term of
$b_f$ is suppressed by the final state fermion mass but it can give
a large contribution in the case of $\phi_\alpha\simeq
2m_{\tilde\chi_\ell^0}$, as is well known.  
In the present models the lightest neutralino can be much lighter than
that of the MSSM. On the other hand, the lightest CP even neutral Higgs scalar 
can be heavier than that of the MSSM as discussed in the previous part.
These features may make the Higgs-pole enhancement effective
in the annihilation of the lightest neutralino.
This is not realized in the nMSSM where the lightest neutral Higgs can
be very light due to the singlet-doublet mixture.  
Since the lightest neutralino can have substantial components which have a new
interaction with the lightest neutral Higgs scalar as shown in
eq.~(\ref{newint}), the Higgs exchange is expected to be important 
if the lightest neutralino contains the ordinary Higgsino or bino component
sufficiently. These aspects have been shown partially by using numerical 
studies in \cite{singlino}. 
In addition to these effects we also have to take account 
of all other processes mediated by the
exchange of the MSSM contents in the numerical estimation of 
the relic abundance of the lightest neutralino because of the following
reasons. Firstly, the lightest neutralino with sufficient
Higgsino components may be light enough. In that case, if 
$m_{\tilde\chi_\ell^0}\simeq m_{Z_1}/2$ and $|U_{14}|^2\not=|U_{15}|^2$
are satisfied as it happens in the S-model \cite{secl},
the annihilation can be enhanced. Secondly,
if the $D$-term contribution of the extra U(1) makes the masses of 
sfermions small enough, the $t$-channel exchange of those sfermions can be a
crucial process for the annihilation of the lightest neutralinos. 
The $D$-term contribution to the sfermion mass is given in 
eq.~(\ref{dterm}) of Appendix. 

\subsection{$Z^\prime$ decay }
Search of $Z^\prime$ is one of important subjects planed at the LHC 
except for the search of Higgs scalars and superpartners 
\cite{lhcevid,lhcevid1}.
Since the present models are characterized by the existence of both the
new neutral gauge boson $Z^\prime$ and the neutral Higgs scalar heavier
than that in the MSSM,\footnote{It is interesting that in a completely 
different context there exist other models which predict both a new neutral 
gauge boson and a neutral Higgs scalar heavier than that expected in 
the MSSM \cite{comph}.} 
the combined analyses of these may give a useful clue for the search of this 
type of models. Here we present some useful formulas for the study 
of the $Z^\prime$ at the LHC.

A tree level cross section for the process 
$pp(p\bar p) \rightarrow Z_2 X\rightarrow f\bar fX$
is given as \cite{lhcevid1}
\begin{equation}
\sigma^f=\sum_q\int^1_0dx_1\int^1_0dx_2~
\sigma(sx_1x_2;q\bar q\rightarrow f\bar f)
G_A^q(x_1,x_2,m_{Z_2}^2)\theta(x_1x_2s-M_\Sigma),
\label{fsigma}
\end{equation}
where $x_{1,2}$ is defined by $x=\sqrt{Q^2\over s}e^y$ 
using the rapidity $y$ and a squared momentum transfer $Q^2$.
The sum of the masses of final state particles is represented 
by $M_\Sigma$. 
$s$ is a square of the center of mass energy in a collision.
A function $G_A^q(x_1,x_2,m_{Z_2}^2)$ depends on the structure 
functions of quarks.
In the present case, eq.~(\ref{fsigma}) can be approximated as 
\cite{lhcevid1}
\begin{equation}
\sigma^f={\kappa\over s}{4\pi^2\over 3}{\Gamma_{Z_2}\over m_{Z_2}}
B(f\bar f)\left[B(u\bar u)+{1\over C_{ud}}B(d\bar d)\right]
C\exp\left(-{\cal A}{M_{Z_2}\over \sqrt{s}}\right),
\end{equation} 
where $C_{ud}=2(25)$, $C=600(300)$ and ${\cal A}=32(20)$ for $pp(p\bar p)$ 
collisions. The QCD correction is taken into account by $\kappa$ and it
is fixed to be $\kappa\sim 1.3$ in the following numerical calculations.
$\Gamma_{Z_2}$ is a total width of $Z_2$ and $B(f\bar f)$ 
is a branching ratio of the $Z_2$ decay into $f\bar f$. 
Formulas for possible decay modes of the $Z_2$ are summarized in Appendix.

We should note that $\sigma^f$ may be expected to take rather different
values from the ordinary ones if the singlino 
dominated lightest neutralino can explain the observed CDM abundance.
Since the decay width into the neutralino sector can be enhanced 
in comparison with the ordinary $Z^\prime $ models, 
the detectability of the $Z_2$ at the LHC may receive a large influence 
as long as the $Z_2$ is searched by using
dilepton events $(f=e,\mu)$ \cite{ce,secl}. We will compare it with the results
obtained in the ordinary $Z^\prime$ models by practicing numerical 
analyses in the next section.

\section{Numerical analyses}
\subsection{Set up for the analyses}
In this section we study parameter space of the models 
allowed by various phenomenological constraints including 
the CDM condition obtained from the analysis of the WMAP data.
Then for such parameters we give some predictions of the models for the
masses of the new neutral gauge field and
the lightest neutral Higgs scalar and the detectability of 
the $Z_2$ at the LHC and so on.

\begin{figure}
\begin{center}
\begin{tabular}{cccc}
   & $Y$  & $Q_\psi$ & $Q_\chi$ \\\hline\hline
$H_1$ & $-1$ & $-2/\sqrt{6}$ & $-2/\sqrt{10}$ \\
$H_2$ & $1$ & $-2/\sqrt{6}$ & $2/\sqrt{10}$ \\
$Q_L$ & $1/3$ & $1/\sqrt{6}$ & $-1/\sqrt{10}$ \\
$L_L$ & $-1$ & $1/\sqrt{6}$ & $3/\sqrt{10}$ \\
\end{tabular}
\end{center}
\vspace*{2mm}
{\footnotesize Table~1~~Abelian charges of the relevant fields 
in a fundamental representation {\bf 27} of $E_6$.}  
\end{figure}

Before proceeding to the results of the analyses, 
we summarize the assumptions which we make in numerical studies.
Firstly, we focus our study on the case that the extra U(1) gaugino
$\tilde\lambda_x$ has a larger mass $M_{\tilde x}$ compared with other
gauginos. This tends to make the lightest neutralino 
dominated by the singlino. 
Secondly, we consider a special case such that 
$\tan\beta\simeq \sqrt{Q_1/Q_2}$ is satisfied so as to relax the
constraint on the value of $u$ and make a phenomenological 
role of the $Z_2$ more effective.
In this case the $ZZ^\prime$ mixing constraint disappears and 
only the constraint derived from the direct search of the $Z_2$ should be 
taken into account.
Thirdly, we restrict our study to the extra U(1)s derived from $E_6$. 
The models with an extra U(1) are constrained from anomaly 
free conditions to be realistic. They are generally required to 
introduce the exotic matter fields to cancel the anomaly. 
If we adopt $E_6$ as a background symmetry, we can control the field 
contents and their U(1)$_x$ charge systematically to make the model 
anomaly free. It is also a promising candidate 
for the extra U(1) models since superstring may realize them as 
its low energy effective theory \cite{efstring}. 

As is well known, $E_6$ has two Abelian factor groups in addition 
to the usual SM gauge group. We assume that only one of 
them remains unbroken at TeV regions and it is broken by 
the VEV $u$ of the SM singlet scalar given in eq.~(\ref{vev}). 
In this case the general U(1)$_x$ can be expressed as a 
linear combination of two representative U(1)s such as \cite{lhcevid1} 
\begin{equation}
Q_x=Q_\psi\cos\theta-Q_\chi\sin\theta,
\label{xcharge}
\end{equation}
where $Q_\psi$ and $Q_\chi$ are the charges of U(1)$_\psi$ and U(1)$_\chi$ 
which are obtained as
\begin{equation}
E_6 \supset SO(10)\times U(1)_\psi, \qquad SO(10)\supset SU(5)\times U(1)_\chi.
\end{equation}
Each charge of the relevant MSSM fields contained in the fundamental
representation {\bf 27} of $E_6$ 
is given in Table 1. The charge of other chiral superfields in the MSSM 
can be determined from them by requiring that the superpotential (\ref{mu}) 
should be invariant under these. Although there are exotic fields such as 
${\bf 3}+{\bf 3}^\ast$ of SU(3) and ${\bf 2}+{\bf 2}^\ast$ of SU(2) 
in {\bf 27} which are not included in the superpotential (\ref{mu}), 
they can be assumed to be sufficiently heavy due to some large VEVs
or the soft supersymmetry breaking effects. Thus, we neglect these
effects on the annihilation of the lightest neutralino
and the $Z_2$ decay in the present analyses. 
Since we consider that U(1)$_x$ is derived from $E_6$, 
the coupling constant of U(1)$_x$ may be related to that of the weak 
hypercharge by $g_x=\sqrt{5\over 3}g_1$, which is derived from
the unification relation 
$${5\over 3}g^2_1\sum_{f\in {\bf 27}}Y_f^2=g_x^2\sum_{f\in {\bf 27}}Q_x^2.$$
We adopt this relation in the present studies. 

In Fig.~1 we plot $\tan\beta$ for the angle $\theta$ which is used in
eq.~(\ref{xcharge}) to define U(1)$_x$. Since we
consider the case of $\tan\beta=\sqrt{Q_1/Q_2}$, $Q_1/Q_2\ge 1$
should be satisfied and then the angle $\theta$ is found to be 
confined into the regions such as $-0.9~{^<_\sim}~\theta~{^<_\sim}~0$.
It is interesting that $\theta=-\tan^{-1}(1/\sqrt{15})\simeq-0.253$ 
is included in this region.
Since the right-handed neutrinos do not have the U(1)$_x$ charge in this
case, they can be very heavy without breaking U(1)$_x$ and the seesaw
mechanism can work to realize the small neutrino mass \cite{extra,kmix1}.
We will study this case in detail as an interesting example 
in the following.

\input epsf 
\begin{figure}[tb]
\begin{center}
\epsfxsize=6.5cm
\leavevmode
\epsfbox{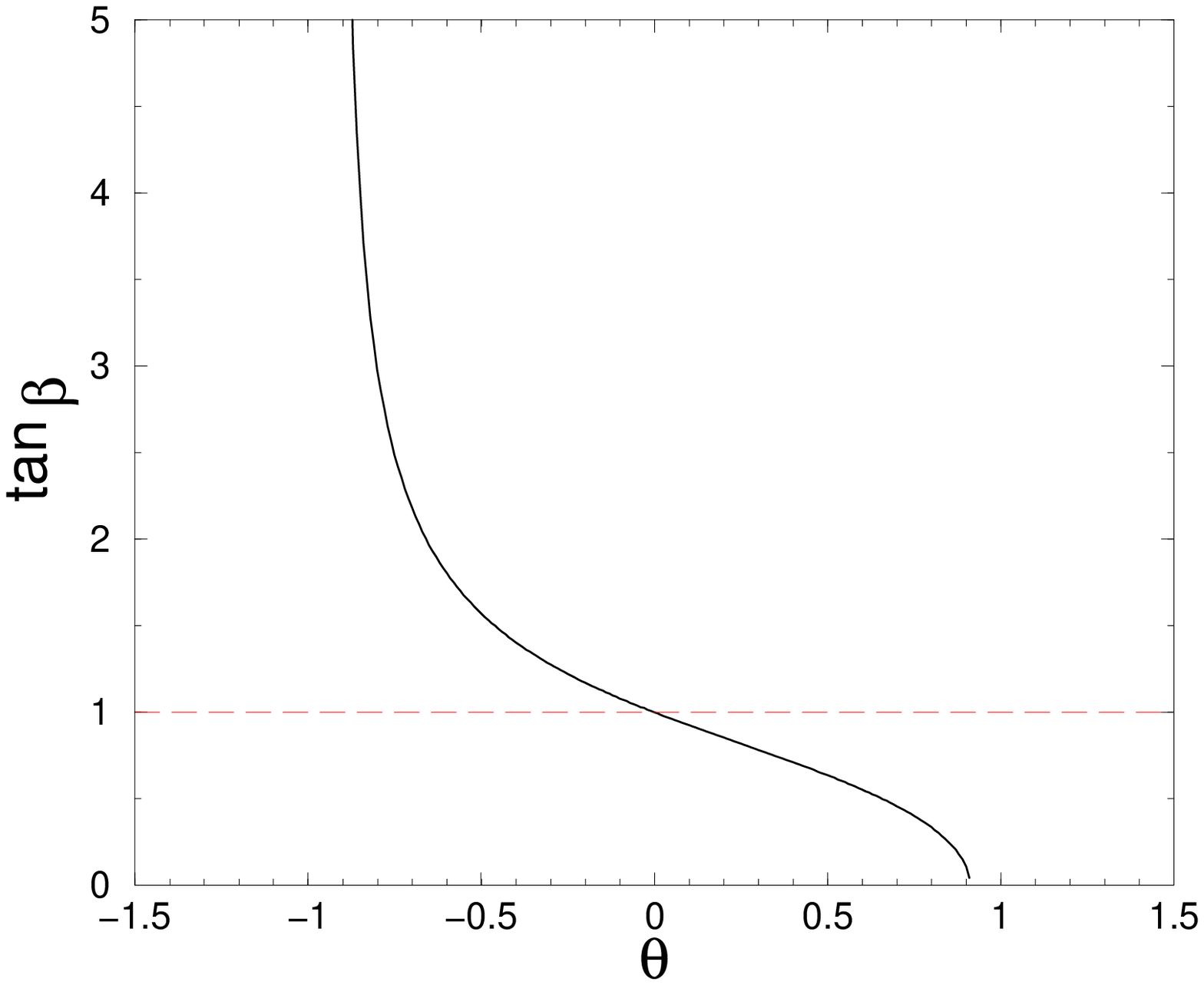}\\
\vspace*{-3mm}
{\footnotesize Fig.~1~~ $\tan\beta$ for various values of $\theta$. }
\end{center}
\end{figure}
   
Now we list up free parameters in the models
for the numerical analyses.
In relation to the extra U(1)$_x$ we have the VEV $u$ and the angle $\theta$.
Since $\tan\beta$ is assumed to be fixed by the U(1)$_x$ charge as
mentioned above,
only $\lambda(\equiv\mu/u )$ is a free parameter in the MSSM sector.
For the soft supersymmetry breaking parameters, we assume the universality 
such as $m_\varphi=m_0$ and $A_U=A_D=\cdots=A$ for the parameters 
in eq.~(\ref{soft}). It can be considered as a result of $E_6$.
Thus, if we assume the unification relation for the masses of gauginos 
$M_{\tilde g}$, $M_{\tilde W}$, $M_{\tilde Y}$ which can be written as  
$M_{\tilde g}=g_3^2M_{\tilde W}/g_2^2$, 
$M_{\tilde Y}=5g_1^2M_{\tilde W}/3g_2^2$, and also 
impose $m_0=A=m_{3/2}$ only to simplify the analyses,
 the number of remaining parameters is reduced into six:\footnote{
It should be noted that this reduction of the model parameters are
caused by the additional assumptions such as $\tan\beta=\sqrt{Q_1/Q_2}$
and $m_0=A$ which are not related to $E_6$.}
$$
\theta, \quad u, \quad \lambda, \quad M_{\tilde W}, 
\quad M_{\tilde x}, \quad m_{3/2}.
$$
We practice the following numerical analyses by scanning the parameters 
$u, M_{\tilde x}$ and $\lambda$ in the following regions:
\begin{eqnarray}
&&300~{\rm GeV}\le u \le 2300~{\rm GeV}~(2~{\rm GeV}), \quad
200~{\rm GeV}\le M_{\tilde x} \le 12~{\rm TeV}~(20~{\rm GeV}), \nonumber \\  
&&200~{\rm GeV}\le M_{\tilde W},~\mu \le 1300~{\rm GeV}~(20~{\rm GeV}),  
\label{para}
\end{eqnarray}
where it should be noted that $\mu$ stands for $\lambda$ for a fixed $u$. 
In the parentheses we show search intervals for these 
parameters.
We fix the supersymmetry breaking parameter $m_{3/2}$ to be $1~{\rm TeV}$ 
as its typical value.\footnote{Since
$M_{\tilde x}$ is assumed to take large values, we need to consider 
its renormalization group effects on other soft supersymmetry breaking 
parameters and also the relation to the radiative symmetry breaking. 
However, since it depends on supersymmetry breaking scenarios, 
we do not go further this subject here. A relevant study can be found in
\cite{mixing}.}

Throughout the analysis  we impose the constraints on the masses 
of the chargino $\tilde\chi^\pm$, sfermions $\tilde f$, 
the lightest CP even neutral Higgs scalar $h^0$ and the charged Higgs
scalar $h^\pm$ as follows: 
\begin{eqnarray}
&&m_{\tilde\chi^\pm}\ge 104~{\rm GeV},\quad  
M_{\tilde g} \ge 195~{\rm GeV}, \quad 
m_{\tilde f}\ge 250~{\rm GeV}, \nonumber\\
&&\quad m_{h^0} \ge 114~{\rm GeV}, \quad m_{h^\pm}\ge 79~{\rm GeV}.
\label{para1}
\end{eqnarray}
Although the constraint on the sfermion masses are model dependent, we
use the bound for the constrained MSSM here as an example. 
The sfermion masses should be checked whether the above
bounds are satisfied by including a $D$-term contribution of the extra U(1).
Since the $D$-term contribution may take a large negative value, 
this condition could give upper bounds for the value of $u$ in that case.
The bounds for $m_{Z_2}$ derived from the direct search of the $Z_2$ are 
taken into account by imposing the Tevatron constraint
\begin{equation}
\sigma(p\bar p\rightarrow Z_2 X)B(Z_2\rightarrow 
e^+e^-,~\mu^+\mu^-)<0.04~{\rm pb}
\label{zpdecay}
\end{equation}
at $\sqrt{s}=1.8$~TeV \cite{cdf}.
We also impose $0\le \lambda\equiv\mu/ u\le 0.75$ which is required by
perturbative bounds for the coupling constant 
$\lambda$ \cite{nmssm1,secl,higgsb}. $\lambda$ is also restricted indirectly 
through the relation to $\mu$ by the chargino mass bound.
The CDM constraint from the 1st year WMAP results is taken into account
as \cite{cmb}
\begin{equation}
0.0945\le \Omega_{\rm CDM}h^2 \le 0.1287 \quad ({\rm at}~2\sigma).
\label{cdm}
\end{equation}
If there exist other dark matter candidates than the lightest neutralino,
the lower bound of (\ref{cdm}) cannot be imposed to constrain the
parameter region. In this paper, however, we make the analysis under the
assumption that the lightest neutralino is the only candidate for the
dark matter.

\subsection{Predictions of the models} 
At first, in order to see global features of the extra U(1) models derived from
$E_6$ we study the $\theta$ dependence of important quantities.
We scan the parameters in the regions shown in (\ref{para}) for each
value of $\theta$ which is allowed from Fig.~1 and search the solutions 
which satisfy all conditions (\ref{para1})--(\ref{cdm}).
In Fig.~2 we plot important physical quantities obtained for these solutions. 
In the left-hand figure we show the predicted regions of the masses of 
the lightest neutralino $\tilde\chi_\ell^0$, 
the lightest neutral Higgs scalar $h^0$ and the new
neutral gauge boson $Z_2$. In the right-hand figure we show the predicted 
cross section for the $Z_2$ decay into the dilepton final 
state $(e^+e^-,\mu^+\mu^-)$ at the LHC with $\sqrt{s}=14$~TeV.
The solutions are found only for more restricted values of $\theta$ 
than those shown in Fig.~1.
Figure 2 suggests that different annihilation processes of the lightest
neutralino play an important role to realize its appropriate relic
density for the CDM in different regions of $\theta$.
The variety of the predicted neutralino mass shows this aspect.

\begin{figure}[tb]
\begin{center}
\epsfxsize=6.5cm
\leavevmode
\epsfbox{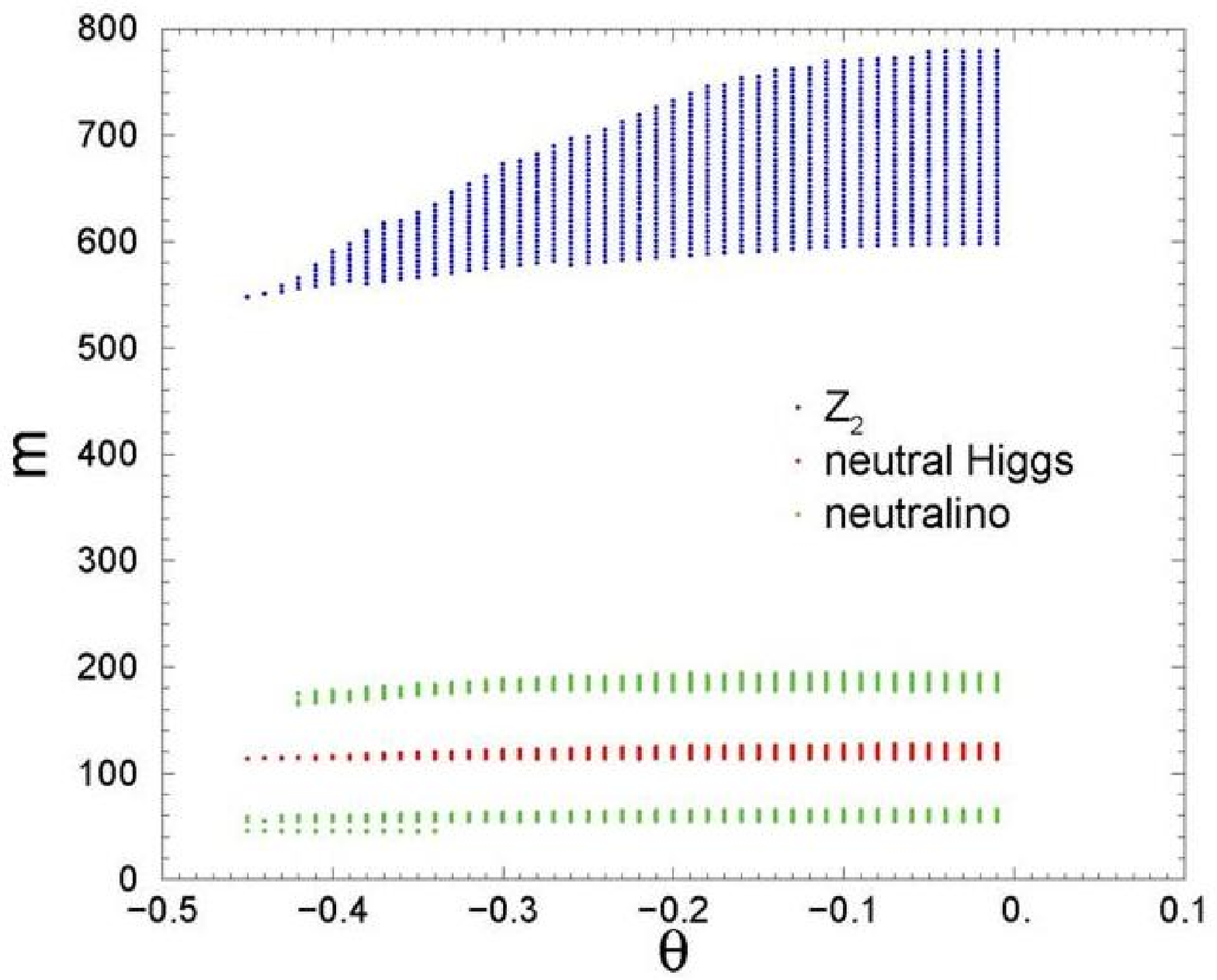}
\hspace*{5mm}
\epsfxsize=6.5cm
\leavevmode
\epsfbox{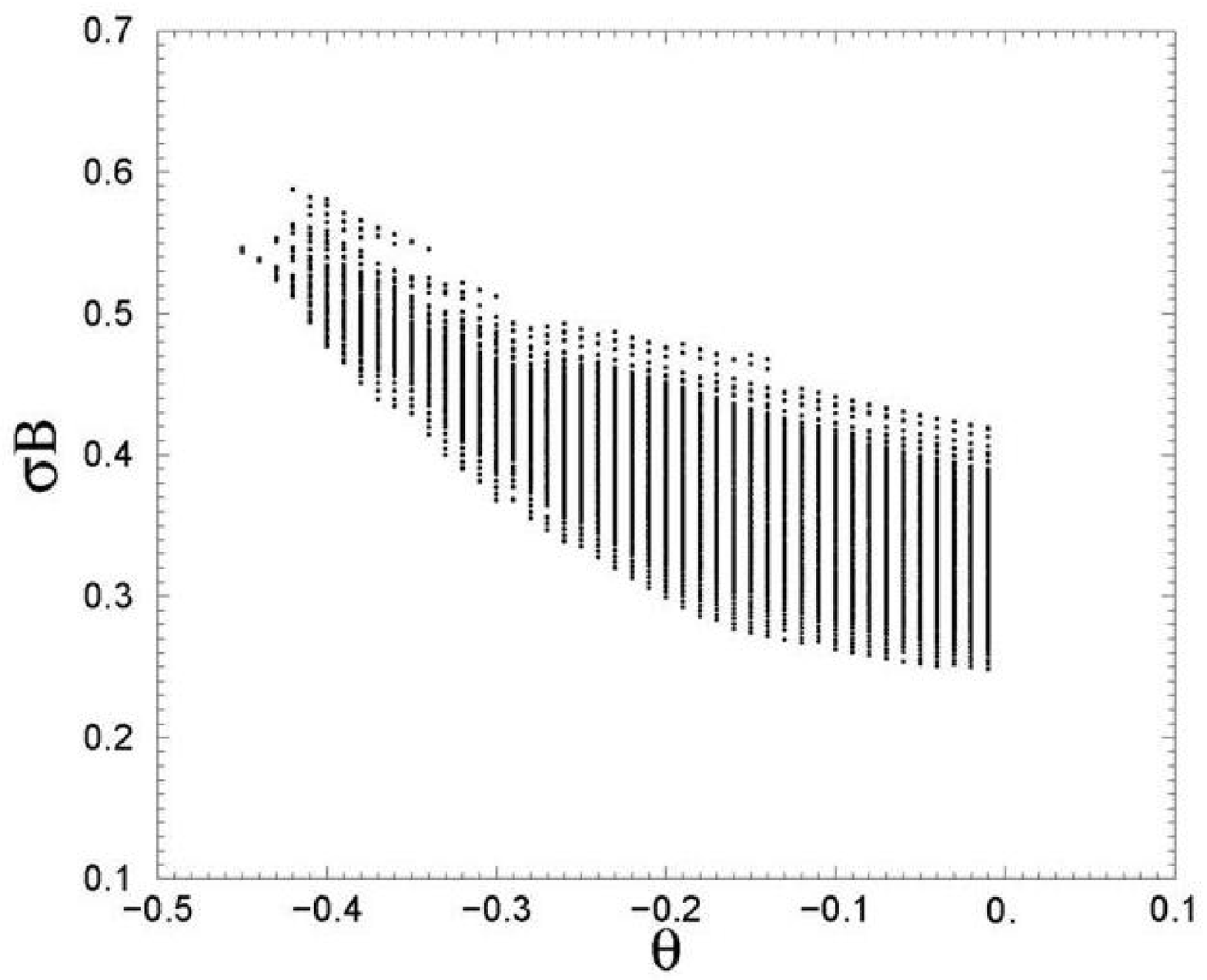}\\
\vspace*{-3mm}
\end{center}
{\footnotesize Fig.~2\ The left-hand figure shows the masses of the 
lightest neutral Higgs, the lightest neutralino and the $Z_2$ 
boson in a GeV unit for various values of $\theta$. 
The right-hand figure shows an expected cross 
section in a fb unit for the $Z_2$ decay into the dilepton final states at 
the LHC with $\sqrt{s}=14$~TeV. These are derived for the parameters
 which satisfy all conditions discussed in the text.}
\end{figure}

As interesting results for $m_{Z_2}$, we find that there exist upper
bounds for $m_{Z_2}$. Since the models considered in the present analysis
have small $\tan\beta$ values, the mass bounds of the lightest neutral
Higgs scalar can give a constraint on $\lambda$ as found from
eq.~(\ref{higgs}). This constraint restricts allowed regions of $u$
through the relation $\mu=\lambda u$. On the other hand, we can find 
the solutions only for restricted regions of $\mu$ around 
650 -- 800 GeV in the present study. 
The upper bounds of $m_{Z_2}$ appear through these backgrounds.
We also find that the lower bound of $m_{Z_2}$ can be less than 600 GeV 
in the allowed regions of $\theta$.
Since the singlino domination of the lightest neutralino makes the decay
width of the $Z_2$ into the neutralino sector larger, the decay width into
the dilepton final state becomes relatively smaller.  Thus 
the constraint imposed by the analysis of the $Z_2$ search at the CDF 
seems to be escapable even for a smaller $m_{Z_2}$ than the usually
discussed values. 
The predicted value of $\sigma B$ shows that $Z_2$ in the present models 
is easily detectable at the LHC.   
Combined searches of the $Z_2$ and the lightest neutral Higgs scalar will be
useful to discriminate the models from other candidates beyond the SM.
    
\subsection{Details of annihilation processes of the lightest neutralino}
As mentioned in the previous part, different annihilation processes seem
to play a crucial role for different regions of $\theta$. 
To clarify these aspects, we examine the models 
defined by typical values of $\theta$ in detail.
We choose $\theta=-0.4$ and $-0.253$ as such examples.   
The features found in these examples are also seen in 
the models defined by other values of $\theta$. 

\begin{figure}[tb]
\begin{center}
\epsfxsize=6.5cm
\leavevmode
\epsfbox{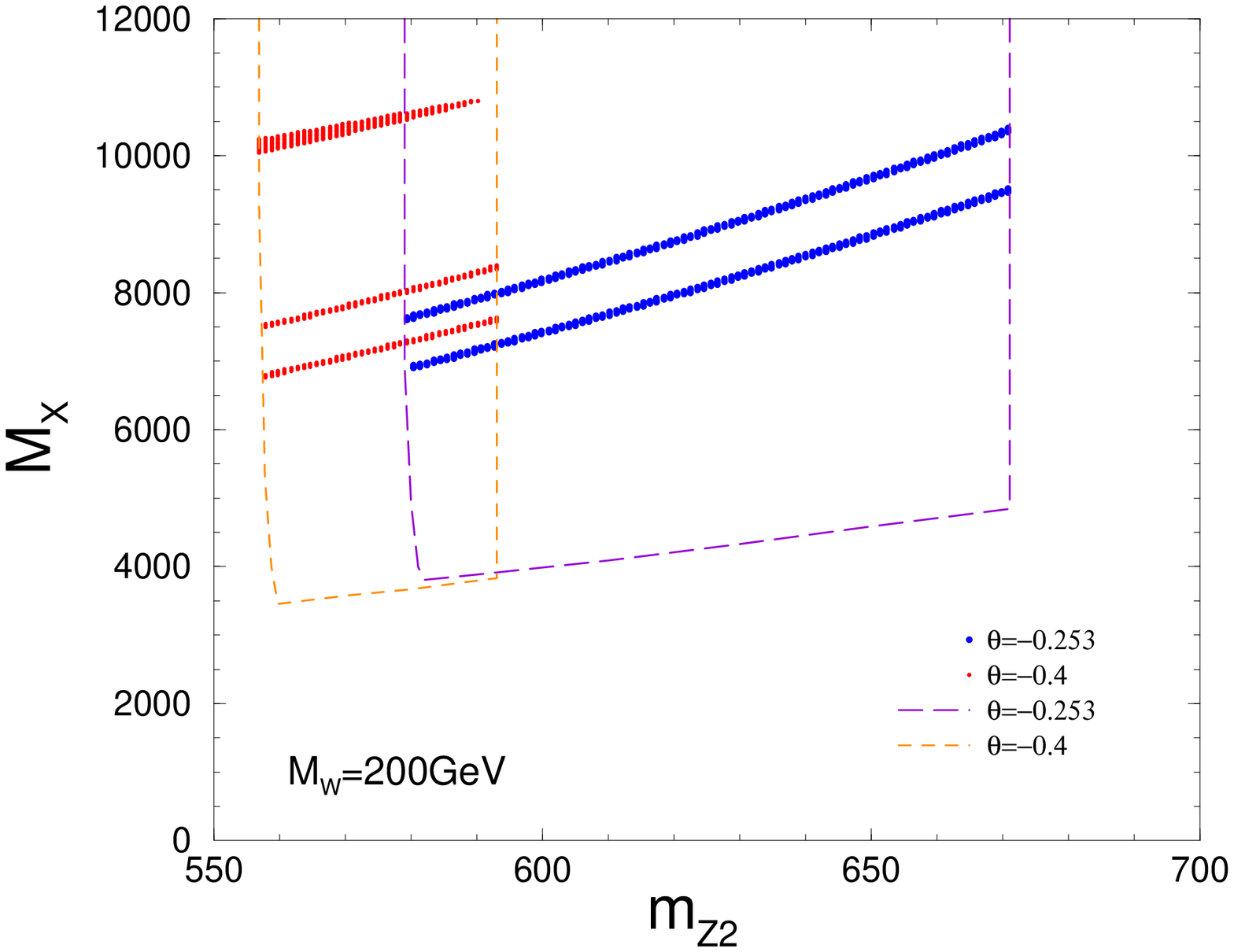}
\hspace*{5mm}
\epsfxsize=6.5cm
\leavevmode
\epsfbox{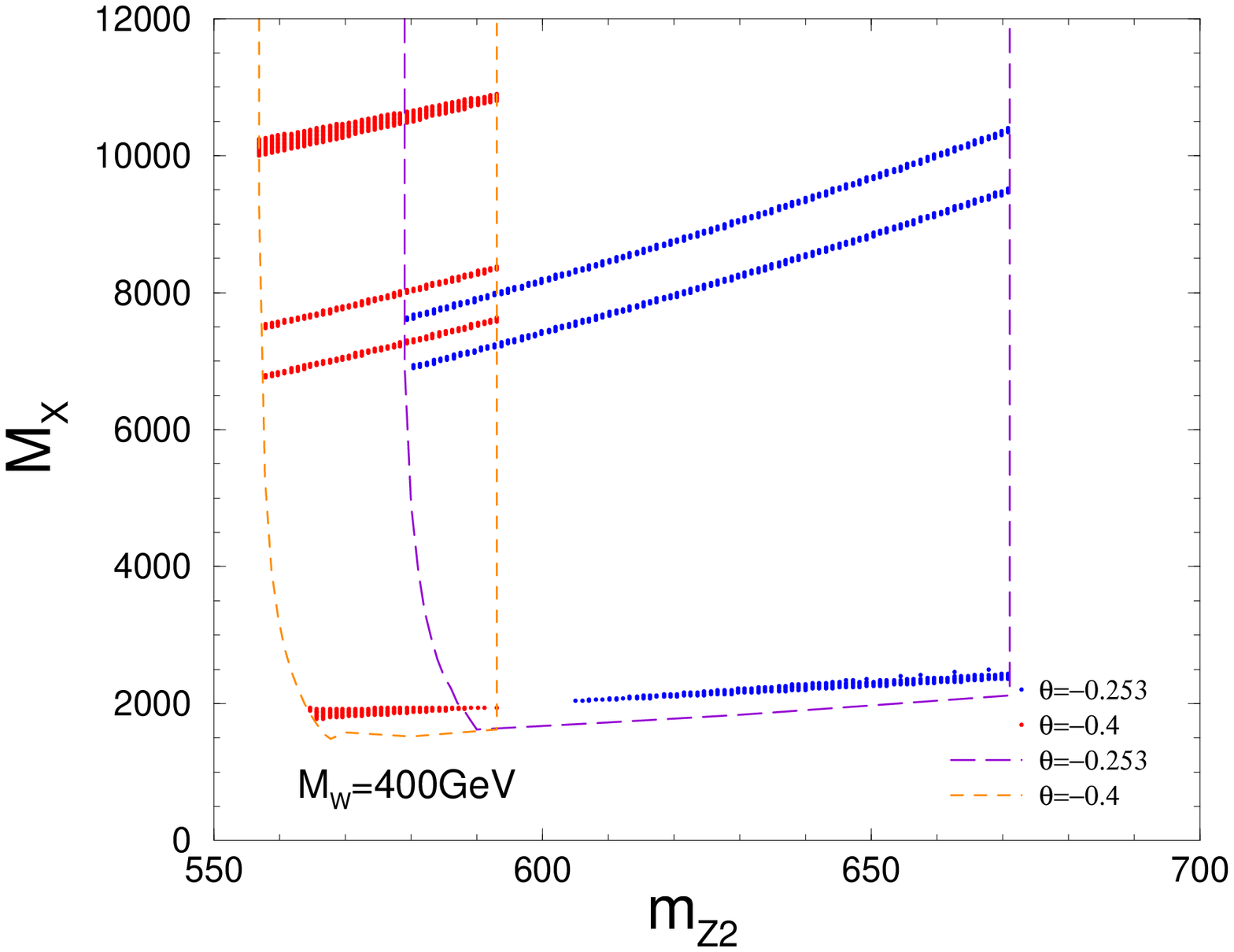}\\
\vspace*{-3mm}
\end{center}
{\footnotesize Fig.~3~~ Regions in the $(m_{Z_2}, M_{\tilde x})$ plane 
which satisfy all conditions discussed in the text.
$\mu=700$~GeV is assumed. 
Values used for $M_{\tilde W}$ and $\theta$ are shown in each figure.  }
\end{figure}

In Fig.~3 we plot points in the $(m_{Z_2},M_{\tilde x})$ plane,
which satisfy all of the required conditions in the cases with 
$\mu=700$ GeV and representative values of $M_{\tilde W}$. 
The used values of $M_{\tilde W}$ are 
shown in the figures. 
In each figure we also show the regions surrounded by a dotted line
($\theta=-0.4$) and a dashed line ($\theta=-0.253$)
where the lightest neutralino is dominated by the singlino $\tilde S$ 
and also the conditions except for (\ref{cdm}) are satisfied. 
The boundary expressed by a vertical line
at a smaller $m_{Z_2}$ is caused by the condition (\ref{zpdecay}) 
which can give the lower bounds of $m_{Z_2}$.
On the other hand, the boundary expressed by a vertical line 
at a larger $m_{Z_2}$ is caused by the mass bounds of the 
lightest neutral Higgs scalar given in (\ref{para1}). 
Since we fix a value of $\mu$, larger values of $u$
make values of $\lambda$ smaller. As we can see from
(\ref{higgs}) and also as discussed in the previous part, 
smaller values of $\lambda$ can decrease the upper bounds of the
Higgs mass and make it below the current experimental bounds 
at small values of $\tan\beta$. 
These features confine the mass of the new gauge boson
into narrow regions as shown in each figure.  
Solutions for other values of $\mu$ have the similar features to those 
shown in this figure except that the solutions for a larger $\mu$ value 
move to the region with larger values of $m_{Z_2}$.

\begin{figure}[tb]
\begin{center}
\epsfxsize=7cm
\leavevmode
\epsfbox{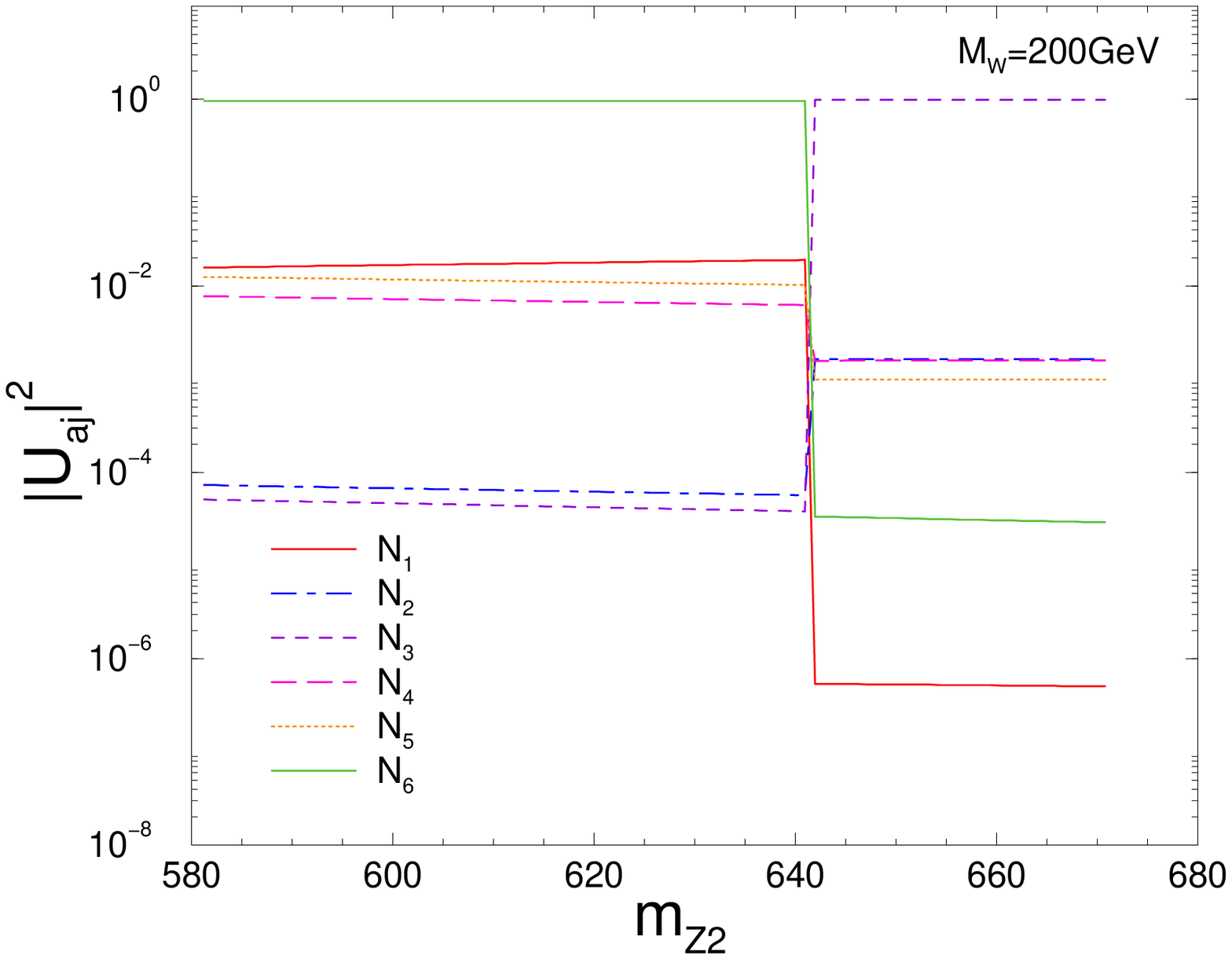}
\vspace*{-3mm}
\end{center}
{\footnotesize Fig.~4~~ Composition of the lightest neutralino in the case of
$\theta=-0.253$, $M_{\tilde x}=4.5$~TeV and $M_{\tilde W}=200$~GeV.}
\end{figure}

From Figs.~3 we find that all solutions are found in the regions
where the lightest neutralino is dominated by the singlino. 
In order to see the change of the composition of the lightest neutralino 
at the lower boundary of the regions of singlino domination, in Fig.~4
we plot values of $|U_{\ell j}|^2$ defined in eq.~(\ref{meig}) 
for $m_{Z_2}$ in the case of $\theta=-0.253$,
$M_{\tilde x}=4.5$~TeV, and $M_{\tilde W}=200$~GeV.
It shows that the singlino domination of the lightest neutralino 
suddenly turns into the bino domination within the narrow region of 
$m_{Z_2}$.  
The other cases with different values of $M_{\tilde W}$ 
and $\mu$ also show the similar behavior to this case on the same boundary.
For larger values of $M_W$ than 400~GeV we find the allowed regions 
in the same places in the $(m_{Z_2},M_{\tilde x})$ plane 
since the composition of the lightest neutralino is almost fixed there.

\begin{figure}[tb]
\begin{center}
\epsfxsize=6.5cm
\leavevmode
\epsfbox{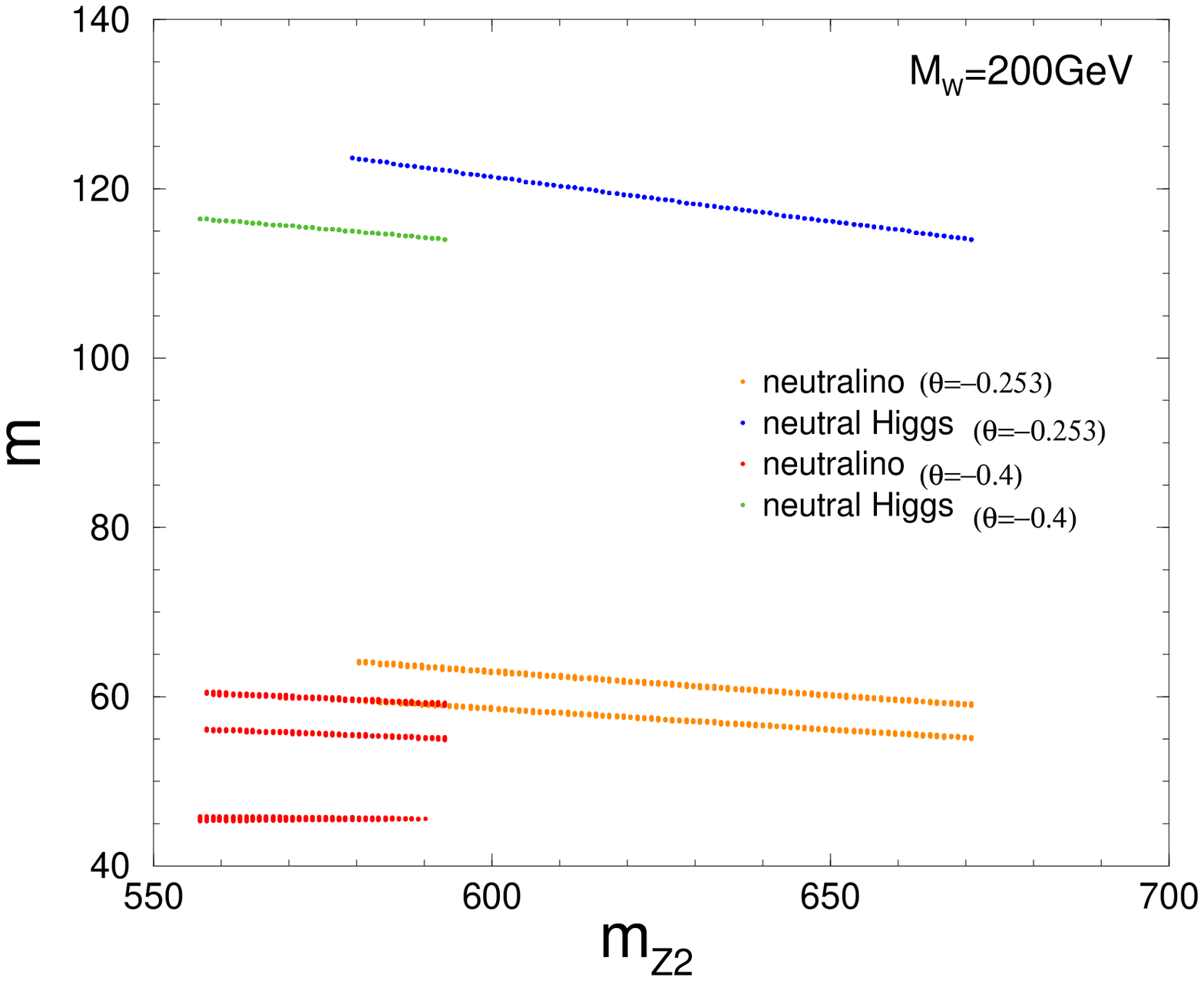}
\hspace*{5mm}
\epsfxsize=6.5cm
\leavevmode
\epsfbox{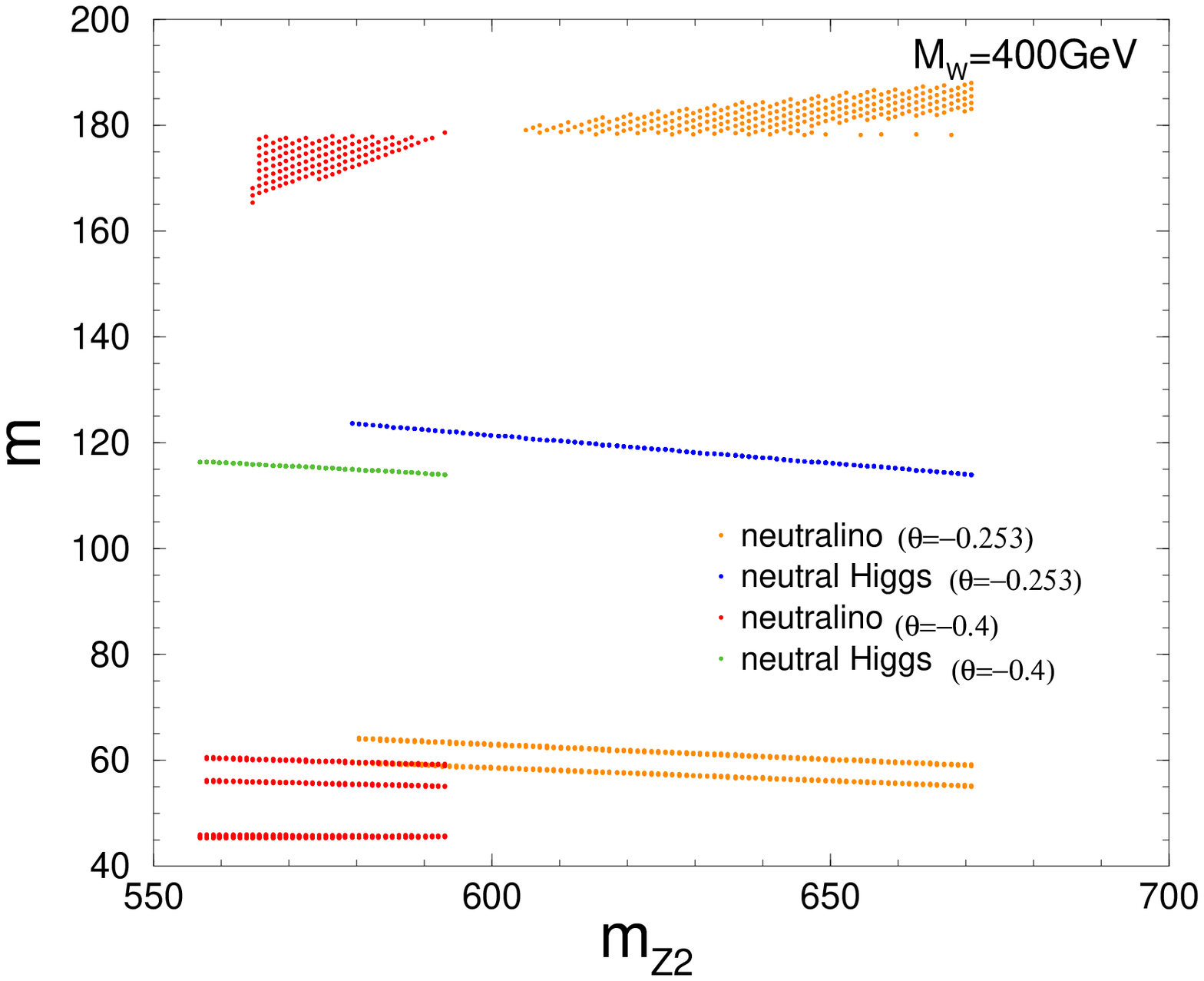}\\
\vspace*{-3mm}
\end{center}
{\footnotesize Fig.~5~~ Masses of the lightest neutral Higgs and the
 lightest neutralino in the cases of $\theta=-0.4$ and $-0.253$. 
$M_{\tilde W}$ is shown in each figure. }
\end{figure}

In Fig.~5 we plot the masses of the lightest neutralino and the
lightest neutral Higgs scalar obtained from the solutions shown in
Fig.~3. Since larger values of $M_{\tilde x}$ produce smaller masses 
for the singlino dominated lightest neutralino as discussed in 
section 2, we can find the correspondence of the solutions 
in Fig.~3 to the mass $m_{\tilde\chi_\ell^0}$ of the 
the lightest neutralino shown in Fig.~5.
We can find from Fig.~5 that there are three types of possibilities for 
the annihilation of the lightest neutralinos to explain the CDM 
abundance. They are 
characterized by the lightest neutralino mass.
The first possibility is characterized by the relatively small 
mass such as $\sim 45$ GeV. It appears commonly 
in the all figures in case of $\theta=-0.4$.
However, we cannot find this type of solutions in case of $\theta=-0.253$.
In these solutions the effective annihilation of the lightest neutralinos 
seems to be induced through a $Z_1$ exchange. 
It can realize the appropriate CDM abundance 
only if $m_{\tilde\chi_\ell^0}\sim m_{Z_1}/2$ and 
the Higgsino components are suitably contained 
in the singlino dominated lightest neutralino.
This conditions seems to be satisfied only in the restricted values of
$\theta$ such as $-0.46~{^<_\sim}~\theta~{^<_\sim}~-0.34$ as found in the
left-hand figure of Fig.~2. 
The second one is characterized by the mass tuned into the
narrow region around $m_{h^0}/2$. These solutions appear as a result
of the Higgs pole enhancement of the annihilation. 
The third one is characterized by the large neutralino masses 
such as $\sim 170-180$ GeV. In this case the annihilation is
considered to be mainly induced by a $Z_2$ exchange. The large mass of the
singlino dominated neutralino can enhance the factor $b_f$ 
in eq.~(\ref{zprime})
to realize the appropriate abundance. This type of solutions appear only
for large values of $M_{\tilde W}$ such as $400$~GeV or more. 
This is because the large $M_{\tilde W}$ is
necessary for the heavy singlino dominated neutralino to be the lightest one. 

\begin{figure}[tb]
\begin{center}
\epsfxsize=7cm
\leavevmode
\epsfbox{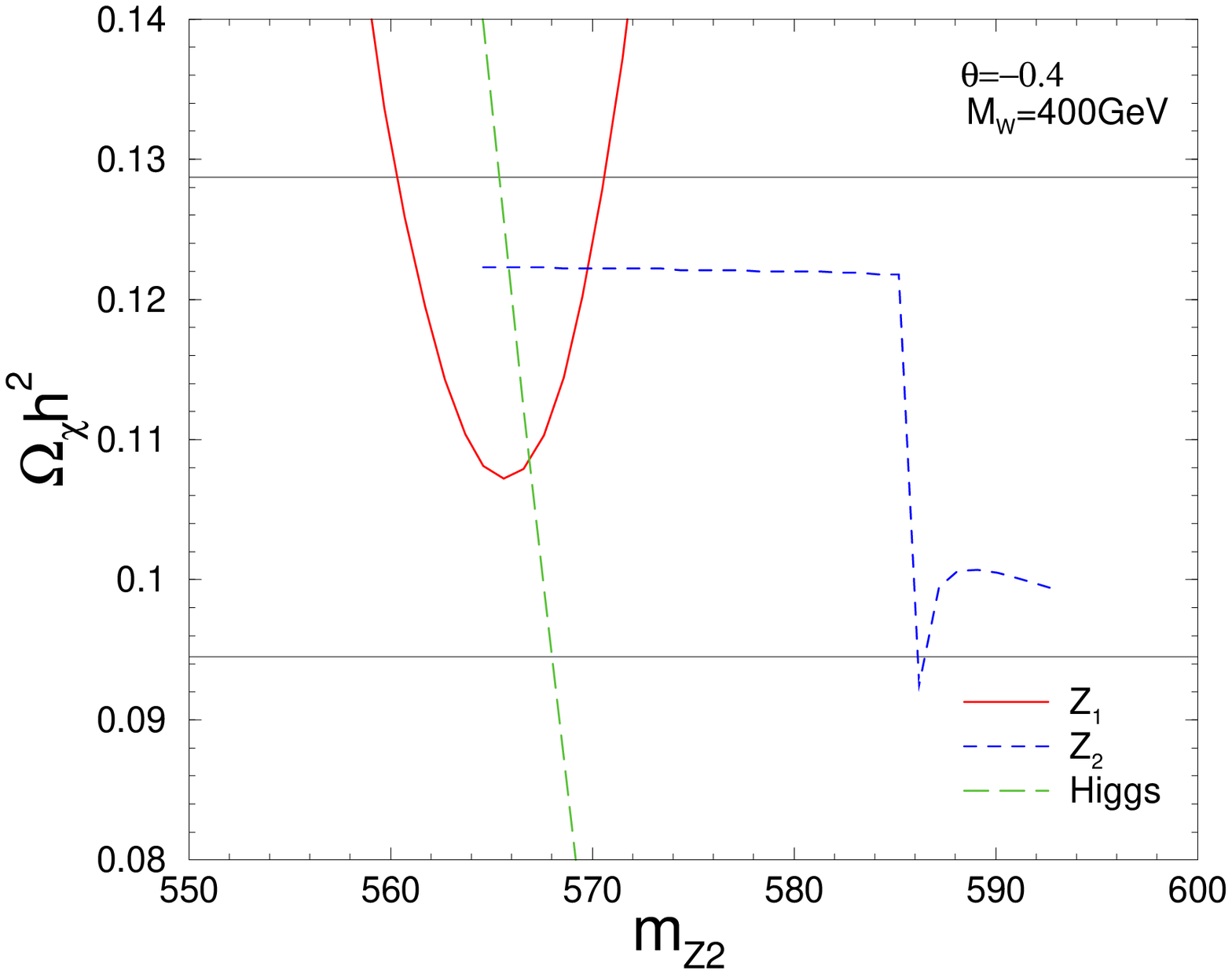}
\end{center}
\vspace*{-3mm}
{\footnotesize Fig.~6~~ Dominant processes which realize the suitable relic 
abundance $\Omega_\chi h^2$ in the case of $\theta=-0.4$ and 
and $M_{\tilde W}=400$~GeV. In order to realize three types of solution 
$M_{\tilde x}$ is chosen such as 10.3 TeV for the $Z_1$ exchange, 7.7~TeV
 for the Higgs exchange, and 1.9~TeV for the $Z_2$ exchange.}
\end{figure}

In order to check these points, we adopt the solutions for $\theta=-0.4$ and
$M_W=400$~GeV in Fig.~3 and calculate $\Omega_\chi h^2$ by taking 
account of only the annihilation process discussed above for each case. 
We choose the values of $M_{\tilde x}$ as 10.3~TeV, 7.7~TeV and 1.9~TeV,
where the three types of solutions appear.
We plot the results in Fig.~6. This figure justifies the above
discussion. Dominant components of the lightest neutralino in each case
are found to be 
\begin{center}
\begin{tabular}{llll}
$Z_1$ exchange ~($M_{\tilde x}=10.3$~TeV)~: & ${\cal N}_6 ~\sim$97.6\%, &  
${\cal N}_5 ~\sim$ 1.4\%,  &\\
$h^0$ exchange ~($M_{\tilde x}=7.7$~TeV)~: & ${\cal N}_6 ~\sim$97.4\%, &  
${\cal N}_5 ~\sim$ 1.4\%, &\\
$Z_2$ exchange ~($M_{\tilde x}=1.9$~TeV)~: & ${\cal N}_6 ~\sim$91\%, &  
${\cal N}_1 ~\sim$7\%,  & ${\cal N}_5 ~\sim$1\%.  \\
\end{tabular} 
\end{center}
For the heavy neutralino solutions the lightest neutralino can be heavier 
than the ordinary $Z_1$. Then we may need to take account of additional new 
final states such as $W^+W^-$ and $Z_1h^0$ which can be mediated by the
$Z_2$ exchange. However, since these processes are generally suppressed
in comparison with the ones with final states $f\bar f$ in the present models, 
these effects seem to be safely neglected in the annihilation of the 
singlino dominated lightest neutralino.
Thus, the present results are considered to be a good approximation 
even in the case of $m_{\tilde\chi_\ell^0}>m_{Z_1}$. 

\begin{figure}[tb]
\begin{center}
\epsfxsize=6.5cm
\leavevmode
\epsfbox{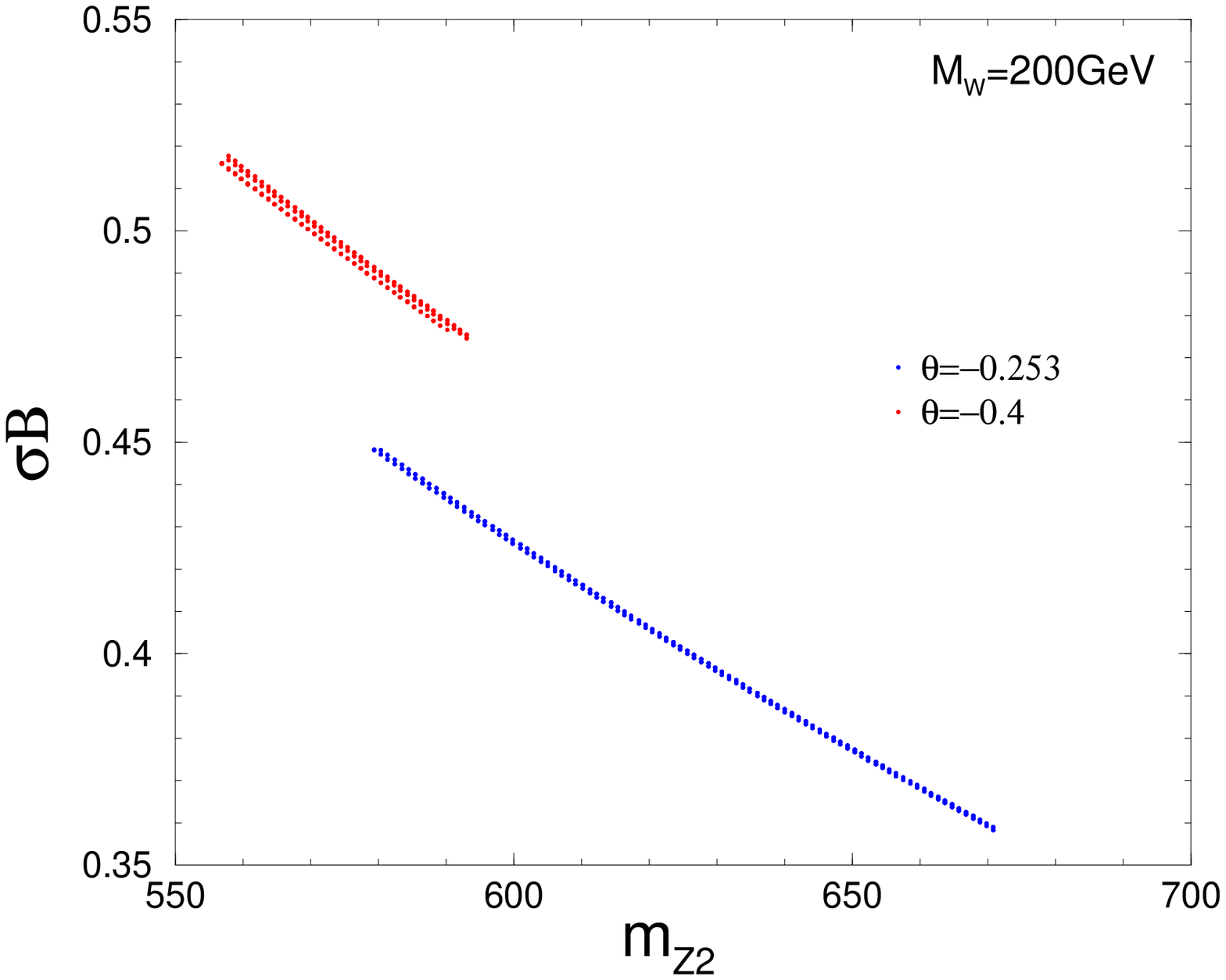}
\hspace*{5mm}
\epsfxsize=6.5cm
\leavevmode
\epsfbox{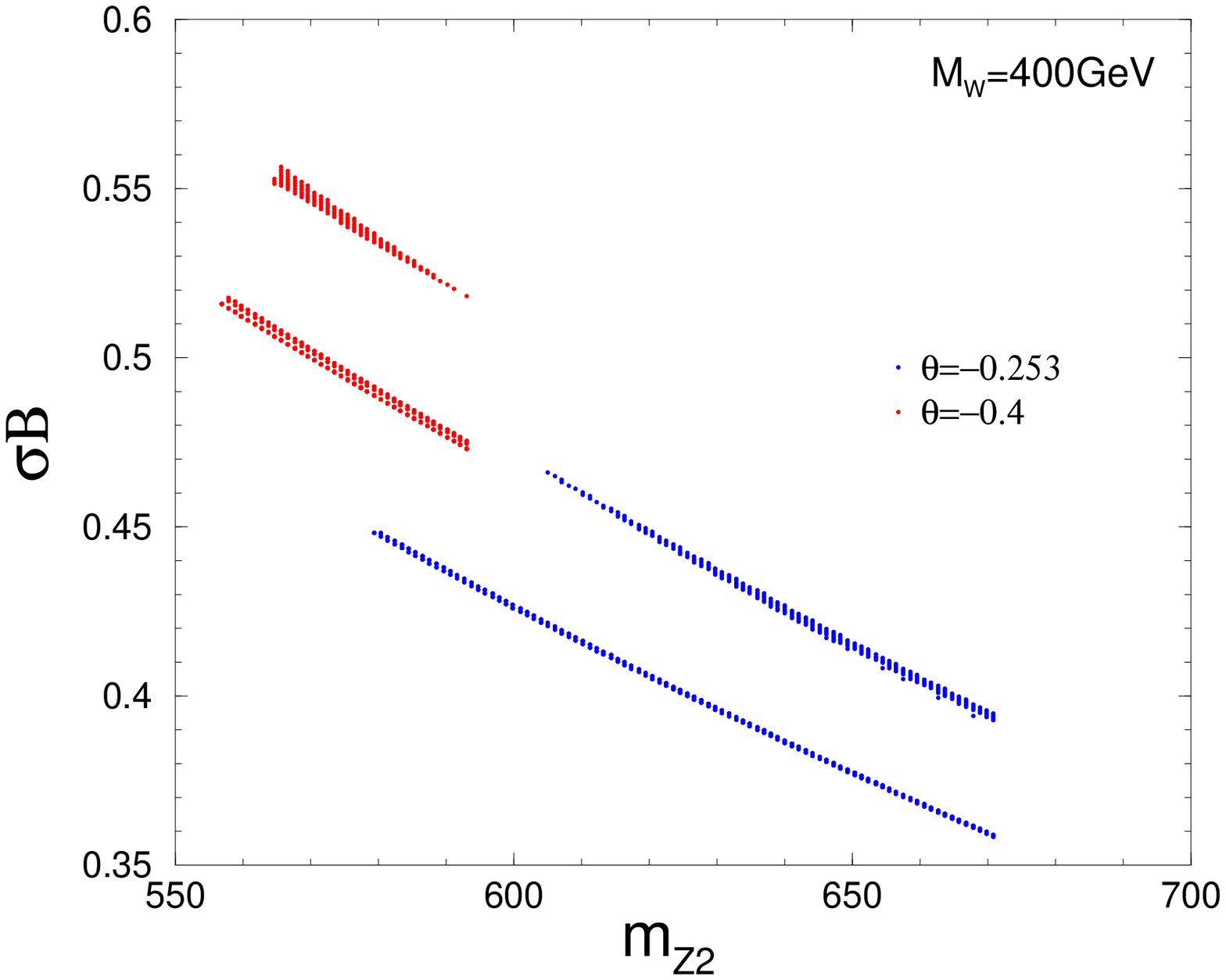}\\
\vspace*{-3mm}
\end{center}
{\footnotesize Fig.~7~~ An expected cross section for the $Z_2$ decay
 into the dilepton final states at the LHC with $\sqrt{s}=14$~TeV in
 the cases of $\theta=-0.4$ and $-0.253$. $\mu=700$~GeV is assumed and 
$M_{\tilde W}$ is shown in each figure. }
\end{figure}
  
In Fig.~7 we plot the predicted cross section $\sigma B$ of the $Z_2$
decay into the dilepton final states $(e^+e^-,\mu^+\mu^-)$ at the LHC with 
$\sqrt{s}=14$~TeV for the models defined by $\theta=-0.4$ and $-0.253$. 
This figure also shows that the LHC can easily find $Z_2$ in the present
models. The models defined by the different values of $\theta$
can also be distinguished from each other.
In the previous part we discussed that the dilepton channel may be suppressed
compared with the case where the lightest neutralino is composed of only
the MSSM
contents, since the $Z_2$ decay into the neutralino pairs can be enhanced in
the case of the singlino dominated lightest neutralino.
This may be generally expected if $\hat S$ has the larger charge of U(1)$_x$
compared with the charged leptons. In the present models this is the case
and the aspect is shown in the figure for $M_{\tilde W}=400$~GeV 
in Fig.~7. 
In the case of $M_{\tilde W}=400$~GeV the appearance of new branches can be
understood from this reasoning. They correspond to the solutions with
the heavier lightest neutralino. Thus the $Z_2$ decay to the neutralino
sector is suppressed and the cross section to the dilepton final states
is enhanced. 

\section{Summary}
We have studied the phenomenology of the models with an extra U(1) 
which can give an elegant weak 
scale solution for the $\mu$ problem and have the singlino dominated 
lightest neutralino.
Such models can be constructed by introducing an SM singlet chiral
superfield $\hat S$ to the MSSM.
The vacuum expectation value $u$ of the scalar component of $\hat S$ generates 
the $\mu$ term and also breaks the extra U(1) symmetry.
The singlino dominated lightest neutralino is realized by assuming 
the extra U(1) gaugino to be heavier than other gauginos.
Under this assumption, the lightest neutralino can be dominated by 
the fermionic component of $\hat S$ even if the vacuum expectation value 
$u$ is large enough to make the extra U(1) gauge field $Z_2$ 
satisfy the current experimental constraints.
We studied the case with $\tan\beta=\sqrt{Q_1/Q_2}$ where the
$ZZ^\prime$ mixing constraint is automatically satisfied even for rather
small value of $u$. Although this fact is known, the study of that case
seems not to be done in the $Z^\prime$ phenomenology. 

The neutralino sector of the models is different 
from the currently known similar type models. 
In the NMSSM and the nMSSM the singlino dominated lightest 
neutralino usually appears for the small values of $u$. 
In the models with an extra U(1) in which the masses of the gauginos
are assumed to be the same order, the lightest neutralino 
is expected to be very similar to that of the MSSM because of a large
value of $u$. On the other hand, in the present models 
the singlino dominated lightest neutralino 
can appear in a very different situation from these and this can make their
phenomenology distinguishable from others.

We have focussed our study on the extra U(1) models which are derived from
 $E_6$. And we have discussed typical phenomenological consequences 
in the cases where the singlino 
dominated lightest neutralino has the suitable relic density 
as a CDM candidate.
We have shown that there are non-negligible parameter
regions which satisfy the currently known phenomenological constraints 
as long as the model satisfies $\tan\beta\simeq\sqrt{Q_1/Q_2}$.
The CDM candidate has been shown to have different natures from that 
in both the MSSM and other type of singlet extensions of the MSSM. 
They may be examined in future collider experiments. 
We have also predicted the masses of the lightest neutral Higgs scalar, the new
neutral gauge boson and the lightest neutralino. 
The predicted cross section for the $Z_2$ decay into the dilepton 
pairs shows that the $Z_2$ boson in the models can be easily 
found at the LHC.
Searches of the neutral Higgs scalar and the $Z_2$ boson at the LHC are 
expected to give us useful informations for the models with 
an extra U(1) which may give the solution for the $\mu$ problem.
  
\vspace*{5mm}
This work is partially supported by a Grant-in-Aid for Scientific
Research (C) from the Japan Society for Promotion of Science (No.17540246).

\newpage
\noindent
{\Large\bf Appendix}

In this appendix we present the interaction Lagrangian relevant 
to possible decay processes of $Z_2$ and also formulas for the 
decay width for them \cite{lhcevid1}.

\noindent
(i)$Z_2 \rightarrow f\bar f$\\
Interaction terms relevant to this decay process are
\begin{eqnarray}
&&{\cal L}={g_x\over 2}\bar f\gamma_\mu(q_v^f-q_a^f\gamma_5)fZ_2^\mu,
\nonumber \\
&&q_v^f={1\over 2}\left(Q(f_L)+Q(f_R)\right), 
\qquad q_a^f={1\over 2}\left(Q(f_L)-Q(f_R)\right).
\label{charge}
\end{eqnarray}
The decay width can be calculated as
\begin{equation}
\Gamma_{f\bar f}=c_f{g_x^2\over 4}{m_{Z_2}\over 12\pi}\left[
q_v^{f2}\left(1+2{m_f^2\over m_{Z_2}^2}\right)
+q_a^{f2}\left(1-4{m_f^2\over m_{Z_2}^2}\right)\right]
\sqrt{1-4{m_f^2\over m_{Z_2}^2}}, 
\end{equation}
where $c_f$ equals to 1 for leptons and 3 for quarks.
Because of various reasons, the extra fermions in the present models 
can be considered 
to be sufficiently heavy and then neglected in the final states
for the $Z_2$ decay. In that case it is justified to confine 
$f$ into ordinary quarks and leptons.

\noindent
(ii)~$Z_2 \rightarrow\tilde f_{L,R}\tilde f_{L,R}^\ast$ \\
Mass matrices of the sfermions have a similar structure to that in the MSSM
except that there are additional contributions from the extra U(1) 
$D$-term. It can play a crucial role since it may give large negative
contributions depending on the models. They can be written as
\begin{equation}
{\cal M}_{\tilde f}^2=\left(\begin{array}{cc}
M_{\tilde f_L\tilde f_L}^2 & M_{\tilde f_L\tilde f_R}^2 \\ 
M_{\tilde f_L\tilde f_R}^{2\dagger} & M_{\tilde f_R\tilde f_R}^2 \\
\end{array}\right),
\label{smatrix}
\end{equation}
where each component of this matrix is a $3\times 3$ matrix with 
respect to flavors. However, if flavor mixing is small enough, 
each flavor can be treated independently and ${\cal M}_{\tilde f}^2$ can be
reduced into a $3\times 3$ matrix. In that case the component of ${\cal
M}_{\tilde f}^2$ can be expressed as
\begin{eqnarray}
&&M_{\tilde f_L\tilde f_L}^2=\tilde m_0^2 +|m_f|^2
+(T_3-Q_fs_W^2)m_Z^2\cos 2\beta+Q_x(f_L)\tilde m_D^2, \nonumber \\
&&M_{\tilde f_R\tilde f_R}^2=\tilde m_0^2 +|m_f|^2+Q_fs_W^2m_Z^2\cos 2\beta
+Q_x(f_R)\tilde m_D^2, \nonumber \\
&&M_{\tilde f_L\tilde f_R}^2=-m_f^\ast(A^\ast-\lambda u R_f),
\end{eqnarray}
where $(T_3, R_f)$ takes $({1\over 2}, \cot\beta)$ and $(-{1\over 2},
\tan\beta)$ for an up- and down-sector of squarks and sleptons, 
respectively. $m_f$ and $Q_f$
are the mass and the electric charge of a fermion $f$.
We assume the universality of soft scalar masses $\tilde m_f^2$
and $A$ parameters.
The $D$-term contribution of the extra U(1) has the expression 
\begin{equation}
\tilde m_D^2={1\over 2}g_x^2\left(Q_1v_1^2+Q_2v_2^2+Q_Su^2\right).
\label{dterm}
\end{equation}
We should note that this contribution may be negative to realize the
light sfermions depending on the value of $\theta$ introduced in
eq.~(\ref{xcharge}) which defines the extra U(1) models.

Interaction terms relevant to this decay process are given by
\begin{eqnarray}
&&{\cal L}={g_x\over 2}(q_v^f\pm q_a^f)\tilde f^\ast_{L,R} 
i\stackrel{\leftrightarrow}{\partial_\mu} f_{L,R}Z_2^\mu,
\end{eqnarray}
where $v_f$ and $a_f$ are defined in eq.(\ref{charge}).
The decay width of this process can be written  as
\begin{equation}
\Gamma_{\tilde f_{L,R}\tilde f_{L,R}^\ast}=c_f{g_x^2\over 4}
{m_{Z_2}\over 48\pi}(q_v^f\pm q_a^f)^2
\left(1 -4{m_{\tilde f_{L,R}}^2\over m_{Z_2}^2}\right)^{3/2}. 
\end{equation}
In this derivation the $LR$ mixing $M_{\tilde f_L\tilde f_R}^2$ 
in eq.~(\ref{smatrix}) is neglected. This is considered to be a good
approximation except for the stop sector.

\noindent
(iii)~$Z_2\rightarrow H^+H^-$\\
The charged Higgs sector has the same structure as that of 
the MSSM except that
there are additional contributions to the mass eigenvalue generated by the
coupling $\lambda \hat S\hat H_1\hat H_2$ as in the NMSSM. 
The mass eigenvalue and its eigenstate can be expressed as
\begin{equation}
m_{H^\pm}^2=m_W^2\left(1-{2\lambda^2\over g_2^2}\right)
+{2A\lambda u\over \sin 2\beta}, \qquad
H^\pm=H_1^\pm\sin\beta+H_2^\pm\cos\beta.
\end{equation}

The interaction Lagrangian relevant to this process is given by
\begin{equation}
{\cal L}={g_x\over 2}(Q_1\sin^2\beta-Q_2\cos^2\beta)
H^+i\stackrel{\leftrightarrow}{\partial_\mu}H^-Z_2^\mu,
\end{equation}
where $G^\pm$ are would-be Goldstone bosons. Using this interaction 
Lagrangian, we can derive the decay width for this process as
\begin{equation}
\Gamma_{H^+H^-}={g_x^2\over 4}{m_{Z_2}\over 48\pi}
(Q_1\sin^2\beta-Q_2\cos^2\beta)^2
\left(1-4{m_{H^\pm}^2\over m_{Z_2}^2}\right)^{3/2}.
\end{equation}

\noindent
(iv)~$Z_2\rightarrow W^\pm H^\mp$\\
The relevant interaction Lagrangian for this process is
\begin{equation}
{\cal L}={g_x\over 2}(Q_1+Q_2)\sin\beta\cos\beta
\left[(m_WH^-W^+_\mu+H^-i\stackrel{\leftrightarrow}{\partial_\mu}G^+)Z_2^\mu
+{\rm h.c.}\right].
\end{equation}
Using this Lagrangian, the decay width for this process can be calculated as
\begin{eqnarray}
\Gamma_{W^\pm H^\mp}&=&{g_x^2\over 4}{m_{Z_2}\over 48\pi}
(Q_1+Q_2)^2\sin^2\beta\cos^2\beta\left[1-2{m_W^2+m_{H^\pm}^2\over m_{Z_2}^2}
+{(m_W^2-m_{H^\pm}^2)^2\over m_{Z_2}^4}\right]^{1/2} \nonumber\\
&\times&\left[1+2{5m_W^2-m_{H^\pm}^2\over m_{Z_2}^2}
+{(m_W^2-m_{H^\pm}^2)^2\over m_{Z_2}^4} \right].
\end{eqnarray}

\noindent
(v)~$Z_2\rightarrow Z_1\phi_\alpha$\\
The relevant interaction Lagrangian is written as
\begin{equation}
{\cal L}=\sum_{\alpha=1}^32{g_x\over 2}m_Z
(Q_1{\cal O}_{1\alpha}\cos\beta -Q_2{\cal O}_{2\alpha}\sin\beta)
(Z_\mu+\partial_\mu G^0)\phi_\alpha Z_2^\mu.
\end{equation}
where we denote a would-be Goldstone boson as $G^0$.
The matrix ${\cal O}$ and the mass eigenvalues $m_{\phi_\alpha}^2$ 
are defined in
eq.~(\ref{hmix}) and also in the text below it. 
The decay width for this process can be derived as
\begin{eqnarray}
\Gamma_{Z\phi_\alpha}&=&\sum_{\alpha=1}^3{g_x^2\over 4}{m_{Z_2}\over 48\pi}
(Q_1{\cal O}_{1\alpha}\cos\beta -Q_2{\cal O}_{2\alpha}\sin\beta)^2\left[ 1
-2{m_{\phi_\alpha}^2+m_{Z_1}^2\over m_{Z_2}^2}+
{(m_{\phi_\alpha}^2-m_{Z_1}^2)^2\over m_{Z_2}^4}\right]^{1/2} \nonumber\\
&\times&\left[1+2{5m_{Z_1}^2-m_{\phi_\alpha}^2\over m_{Z_2}^2}+
{(m_{\phi_\alpha}^2-m_{Z_1}^2)^2\over m_{Z_2}^4}\right].
\end{eqnarray}

\noindent
(vi)~$Z_2\rightarrow \phi_\alpha P_A$\\
$P_A$ is the CP odd Higgs scalar.
The relevant interaction Lagrangian for this process is written as
\begin{equation}
{\cal L}=\sum_{\alpha=1}^3{g_x\over 2}
\left(Q_1{\cal O}_{1\alpha}{u\sin\beta\over N}
+Q_2{\cal O}_{2\alpha}{u\cos\beta\over N}
+Q_S{\cal O}_{3\alpha}{v\sin\beta\cos\beta\over N}\right)Z_2^\mu
P_A\stackrel{\leftrightarrow}{\partial_\mu}\phi_\alpha,
\end{equation}
where $N$ is defined in eq.~(\ref{cpodd}) and in the text below it.
Using this, we can derive the decay width as
\begin{eqnarray}
\Gamma_{\phi_\alpha P_A}&=&{g_x^2\over 4}{m_{Z_2}\over 48\pi}{1\over N^2}
(Q_1{\cal O}_{1\alpha}u\sin\beta+Q_2{\cal O}_{2\alpha}u\cos\beta
+Q_S{\cal O}_{3\alpha}v\sin\beta\cos\beta)^2 \nonumber \\
&\times& \left[ 1-2{m_{\phi_\alpha}^2+m_{P_A}^2\over m_{Z_2}^2}+
{(m_{\phi_\alpha}^2-m_{P_A}^2)^2\over m_{Z_2}^4}\right]^{3/2}.
\end{eqnarray}

\noindent
(vii)~$Z_2\rightarrow \tilde \chi_a^0\tilde \chi_b^0$\\
The neutralino sector is discussed in detail in the text.
The interaction Lagrangian for the neutralinos $\tilde\chi_a^0$ 
and $Z_2$ is given by
\begin{equation}
{\cal L}=\sum_{a,b=1}^6 g_{ab}\bar{\tilde \chi_a^0}
\gamma_\mu\gamma_5 \tilde \chi_b^0Z_2^\mu, 
\end{equation}
where an effective coupling $g_{ab}$ is defined as
\begin{equation}
g_{ab}=\sum_{j=4}^6{g_x\over 2}Q({\cal N}_j)U_{aj}U^\ast_{bj}.
\end{equation}
The definition of the matrix $U$ is given in eq.~(\ref{meig}).
Using this interaction Lagrangian, the decay width is derived as
\begin{eqnarray}
\Gamma_{\tilde\chi_a^0\tilde\chi_b^0}&=&{g_{ab}^2m_{Z_2}\over 12\pi}
\left(1-{1\over 2}\delta_{ab}\right)
\left[1-{m^2_a+m_b^2\over 2m_{Z_2}^2}
-{(m_a^2-m_b^2)^2\over 2m_{Z_2}^4}-{3m_am_b\over m_{Z_2}^2}\right] \nonumber\\
&\times&\left[1-2{m_a^2+m_b^2\over m_{Z_2}^2}
+{(m_a^2-m_b^2)^2\over m_{Z_2}^4}\right]^{1/2},
\end{eqnarray}
where $m_a$ stands for the mass eigenvalue of the neutralino $\tilde\chi_a^0$.

\noindent
(viii)~$Z_2\rightarrow \tilde C_a^+\tilde C_b^-$\\
The chargino sector has the same structure as that of the MSSM.
The mass terms can be expressed as
\begin{equation}
{1\over 2}(-i\tilde\lambda_{W^-},\tilde H_1^-)
\left(\begin{array}{cc}M_{\tilde W} & \sqrt{2}m_W\sin\beta \\
\sqrt{2}m_W\cos\beta & -\lambda u\\\end{array}\right)
\left(\begin{array}{c}
-i\tilde\lambda_{W^+}\\ \tilde H_2^+\\\end{array}\right)+ {\rm h.c.}.
\end{equation}
If we define the mass eigenstates as
\begin{eqnarray}
&& \left(\begin{array}{c}\psi_1^+ \\ \psi_2^+ \\
\end{array}\right)
=V_+\left(\begin{array}{c}-i\tilde\lambda_{W^+} \\ \tilde H_2^+ \\
\end{array}\right), \qquad 
V_+\equiv\left(\begin{array}{cc}\cos\phi_+ &-\sin\phi_+\\
\sin\phi_+ & \cos\phi_+ \\\end{array}\right), \nonumber \\
&& \left(\begin{array}{c}\chi_2^- \\ \chi_1^- \\
\end{array}\right)
=V_-\left(\begin{array}{c}-i\tilde\lambda_{W^-} \\ \tilde H_1^- \\
\end{array}\right), \qquad 
V_-\equiv\left(\begin{array}{cc}\cos\phi_- &-\sin\phi_-\\
\sin\phi_- & \cos\phi_- \\\end{array}\right),
\end{eqnarray}
the  mixing angles $\phi_\pm$ have analytic expressions 
\begin{eqnarray}
&&\tan 2\phi_+={2\sqrt{2}m_W(M_{\tilde W}\sin\beta-\lambda u\cos\beta)\over 
2m_W^2(\sin^2\beta-\cos^2\beta)+\lambda^2u^2-M_{\tilde W}^2}, \nonumber\\
&&\tan 2\phi_-={2\sqrt{2}m_W(M_{\tilde W}\cos\beta-\lambda u\sin\beta)\over 
2m_W^2(\cos^2\beta-\sin^2\beta)+\lambda^2u^2-M_{\tilde W}^2}.
\end{eqnarray}
Using this basis, the mass terms can be transformed into
\begin{equation}
{1\over 2}(\chi_1^-, \chi_2^-)\left(\begin{array}{cc}
m_1 & 0\\ 0 &m_2 \\ \end{array}\right)\left(\begin{array}{c}
\psi_1^+ \\ \psi_2^+\\
\end{array}\right) + {\rm h.c.}, 
\end{equation}
where the mass eigenvalues are represented as
\begin{eqnarray}
&&m_1=(M_{\tilde W}\sin\phi_-+\sqrt{2}m_W\cos\beta\cos\phi_-)\cos\phi_+ 
\nonumber\\
&&\hspace*{40mm}-(\sqrt{2}m_W\sin\beta\sin\phi_--\lambda u\cos\phi_-)
\sin\phi_+, \nonumber\\
&&m_2=(M_{\tilde W}\cos\phi_- -\sqrt{2}m_W\cos\beta\sin\phi_-)\sin\phi_+ 
\nonumber\\
&&\hspace*{40mm}+(\sqrt{2}m_W\sin\beta\cos\phi_- 
+\lambda u\sin\phi_-)\cos\phi_+.
\end{eqnarray}
Since only the Higgsinos $\tilde H_{1,2}$ interact with $Z_2$,
the interaction Lagrangian relevant to the $Z_2$ decay is given by
\begin{equation}
{\cal L}={g_x\over 2}\sum_{a,b=1}^2
\bar{\tilde C_a}\gamma_\mu(v_{ab}+a_{ab}\gamma_5)\tilde C_bZ_2^\mu, 
\end{equation}
where $\tilde C_a^{\pm T}=(\psi_a^\pm, \chi_a^\pm)$, and 
effective couplings $v_{ab}$ and $a_{ab}$ are defined as
\begin{eqnarray}
&&v_{11}=-{Q_1\over 2}\sin^2\phi_-+{Q_2\over 2}\sin^2\phi_+, \qquad
a_{11}={Q_1\over 2}\sin^2\phi_-+{Q_2\over 2}\sin^2\phi_+, \nonumber \\
&&v_{22}=-{Q_1\over 2}\cos^2\phi_-+{Q_2\over 2}\cos^2\phi_+, \qquad
a_{22}={Q_1\over 2}\cos^2\phi_-+{Q_2\over 2}\cos^2\phi_+, \nonumber \\
&& v_{12}=v_{21}={Q_1\over 2}\sin\phi_-\cos\phi_-
-{Q_2\over 2}\sin\phi_+\cos\phi_+, \nonumber\\
&&a_{12}=a_{21}=-{Q_1\over 2}\sin\phi_-\cos\phi_-
-{Q_2\over 2}\sin\phi_+\cos\phi_+.
\end{eqnarray}
Using these couplings, the decay width for this process is derived as
\begin{eqnarray}
\Gamma_{\tilde C_a\tilde C_b}&=&{g_x^2\over 4}{m_{Z_2}\over 48\pi}\left[
(v_{ab}^2+a_{ab}^2)\left(1-{m^2_a+m_b^2\over 2m_{Z_2}^2}
-{(m_a^2-m_b^2)^2\over 2m_{Z_2}^4}\right)+3(v_{ab}^2-a_{ab}^2)
{m_am_b\over m_{Z_2}^2}\right] \nonumber\\
&\times&\left[1-2{m_a^2+m_b^2\over m_{Z_2}^2}
+{(m_a^2-m_b^2)^2\over m_{Z_2}^4}\right]^{1/2}.
\end{eqnarray}

Finally, the branching ratio $B(Z_2\rightarrow XY)$ used in the text 
is defined by
\begin{equation}
B(Z_2\rightarrow XY)={\Gamma(Z_2\rightarrow XY) \over \Gamma_{\rm tot}},
\end{equation} 
where the total decay width $\Gamma_{\rm tot}$ of $Z_2$ is
\begin{eqnarray}
\Gamma_{\rm tot}&=&\sum_f\Gamma_{f\bar f}
+\sum_{\tilde f}\Gamma_{\tilde f_{L,R}\tilde f_{L,R}^\ast}
+\Gamma_{H^+H^-}+\Gamma_{W^\pm H^\mp} \nonumber \\
&+&\sum_{i=1}^3\Gamma_{Z\phi_i}
+\sum_{j=1}^3\Gamma_{\phi_jP_A}
+\sum_{a,b=1}^6\Gamma_{\tilde N_a\tilde N_b}
+\sum_{a,b=1}^2\Gamma_{\tilde C_a\tilde C_b}.
\end{eqnarray}


\end{document}